\documentclass[pdflatex,sn-mathphys-num,iicol]{sn-jnl}

\usepackage{lmodern}
\usepackage{graphicx}%
\usepackage{multirow}%
\usepackage{amsmath,amssymb,amsfonts}%
\usepackage{amsthm}%
\usepackage{mathrsfs}%
\usepackage[title]{appendix}%
\usepackage{xcolor}%
\usepackage{color}
\usepackage{soul}
\usepackage{textcomp}%
\usepackage{manyfoot}%
\usepackage{booktabs}%
\usepackage{algorithm}%
\usepackage{algorithmicx}%
\usepackage{algpseudocode}%
\usepackage{listings}%
\usepackage[serbian, english]{babel}
\usepackage[normalem]{ulem}
\usepackage{caption}
\usepackage{subcaption}
\usepackage{tabularx}
\usepackage{siunitx}
\usepackage{bm}
\usepackage[stretch=10]{microtype}
\usepackage{hyperref}
\hypersetup{
    colorlinks = true,
    linkcolor = blue,
    anchorcolor = blue,
    citecolor = blue,
    filecolor = blue,
    urlcolor = blue
    }
\usepackage{upgreek}
\usepackage{mathtools}
\usepackage{relsize}
\usepackage{import}
\usepackage{xspace}

\DeclareSIUnit{\magneton}{\ensuremath{\mu_{\mathrm{N}}}}
\DeclareSIUnit{\eVolt}{\ensuremath{\mathrm{eV}}}
\DeclareSIUnit{\bar}{\ensuremath{\mathrm{bar}}}

\theoremstyle{thmstyleone}%
\theoremstyle{thmstyletwo}%
\theoremstyle{thmstylethree}%
\begin{document}
\title[Laser spectroscopy in Tm isotopes]{Laser Spectroscopy of Thulium Isotopes Near the $N=82$ Shell Closure: Nuclear Magnetic Dipole Moment and Charge Radius of ${}^{152\mathrm{m}}\mathrm{Tm}$}



\author*[1,2,3]{\fnm{Jana} \sur{Weyrich}}\email{j.weyrich@gsi.de}
\equalcont{These authors contributed equally to this work. It will be part of the dissertation of Jana Weyrich and Kenneth van Beek.}

\author*[1,4]{\fnm{Kenneth} \spfx{van} \sur{Beek}}\email{k.vanbeek@gsi.de}
\equalcont{These authors contributed equally to this work. It will be part of the dissertation of Jana Weyrich and Kenneth van Beek.}

\author[5]{\fnm{Harshithbabu}}
\author[2,3]{\fnm{Aayush} \sur{Arya}}
\author[3]{\fnm{Sebastian} \sur{Berndt}}
\author[1,2,3]{\fnm{Michael} \sur{Block}}
\author[1,6]{\fnm{Alexandre} \sur{Brizard}}
\author[1,2,3]{\fnm{Premaditya} \sur{Chhetri}}
\author[5,§]{\fnm{Arno} \sur{Claessens}}
\author[1,2,3]{\fnm{Christoph Emanuel}\sur{Düllmann}}
\author[5]{\fnm{Rafael} \sur{Ferrer}}
\author[6]{\fnm{Sarina} \sur{Geldhof}}
\author[1,2]{\fnm{Francesca} \sur{Giacoppo}}
\author[1,2,7]{\fnm{Manuel J.} \sur{Gutiérrez}}
\author[3]{\fnm{Raphael} \sur{Hasse}}
\author[2,3]{\fnm{Christian} \sur{Helmel}}
\author[1]{\fnm{Fritz Peter} \sur{Heßberger}}
\author[1,2,3]{\fnm{Julian} \sur{Hindermann}}
\author[5]{\fnm{Fedor} \sur{Ivandikov}}
\author[2,3]{\fnm{Biswajit} \sur{Jana}}
\author[1,2]{\fnm{Tom} \sur{Kieck}}
\author[6]{\fnm{Mustapha} \sur{Laatiaoui}}
\author[6]{\fnm{Nathalie} \sur{Lecesne}}
\author[1,2]{\fnm{Andrew} \sur{Mistry}}
\author[1,2,3]{\fnm{Danny} \sur{Münzberg}}
\author[3]{\fnm{Thorben} \sur{Niemeyer}}
\author[1,2]{\fnm{Sebastian} \sur{Raeder}}
\author[1,2,3]{\fnm{Elisabeth} \sur{Rickert}}
\author[9]{\fnm{Daniel} \sur{Rodríguez}}
\author[6]{\fnm{Hervé} \sur{Savajols}}
\author[3]{\fnm{Matou} \sur{Stemmler}}
\author[1,2]{\fnm{Dominik} \sur{Studer}}
\author[4]{\fnm{Thomas} \sur{Walther}}
\author[10]{\fnm{Jessica} \sur{Warbinek}}
\author[5]{\fnm{Piet} \spfx{Van} \sur{Duppen}}
\author[3]{\fnm{Klaus} \sur{Wendt}}



\affil[1]{\orgname{GSI Helmholtzzentrum für Schwerionenforschung GmbH}, \orgaddress{\street{Planckstraße 1}, \city{Darmstadt}, \postcode{64291}, \country{Germany}}}

\affil[2]{\orgname{Helmholtz-Institut Mainz}, \orgaddress{\street{Staudingerweg 18}, \city{Mainz}, \postcode{55128}, \country{Germany}}}

\affil[3]{\orgname{Johannes Gutenberg-Universität Mainz}, \city{Mainz}, \postcode{55099}, \country{Germany}}

\affil[4]{\orgname{Technische Universität Darmstadt}, \orgaddress{\street{Schlossgartenstrasse 7}, \city{Darmstadt}, \postcode{64289}, \country{Germany}}}

\affil[5]{\orgname{KU Leuven}, \orgaddress{\street{Celestijnenlaan 200D}, \city{Leuven}, \postcode{B-3001, 3000}, \country{Belgium}}}

\affil[6]{\orgname{Grand Accelerateur National d’Ions Lourds}, \street{Bd Henri Becquerel}, \city{Caen}, \postcode{BP 55027—14076 Caen Cedex 05}, \country{France}}

\affil[7]{\orgname{Universität Greifswald}, \city{Greifswald}, \street{Domstraße 11}, \postcode{17489}, \country{Germany}}

\affil[8]{\orgname{Universidad de Granada}, \street{Avda. Fuentenueva s/n}, \city{Granada}, \postcode{18071}, \country{Spain}}

\affil[9]{\orgname{CERN}, \street{Esplanade des Particules 1}, \city{Geneva}, \postcode{1211}, \country{Switzerland}}

\affil[§]{\orgname{Current Address: Facility for Rare Isotope Beams, Michigan State University}, \street{640 South Shaw Lane}, \city{East Lansing}, \postcode{48824}, \country{USA}}




\abstract{We report on resonance ionization laser spectroscopy measurements performed on both neutron-deficient and neutron-rich thulium (\unboldmath$\mathrm{Tm}, Z=69$) isotopes.
Isotope shifts were determined for three atomic ground-state transitions at wavelengths of \SI{388.4}{\nano\meter}, \SI{388.8}{\nano\meter}, and \SI{389.8}{\nano\meter} in the isotopes \unboldmath${}^{152\mathrm{m}}\mathrm{Tm}$, \unboldmath${}^{153}\mathrm{Tm}$, ${}^{154\mathrm{m}}\mathrm{Tm}$, and \unboldmath${}^{169}\mathrm{Tm}$.
For the \SI{389.8}{\nano\meter} transition, measurements were extended to \unboldmath${}^{170}\mathrm{Tm}$. The hyperfine structures were partially resolved for all five isotopes. For this transition, the isotope shift could be determined for one more isotope, \unboldmath${}^{154}\mathrm{Tm}$.
From the extracted hyperfine coupling constants, the nuclear magnetic dipole moment for \unboldmath${}^{152\mathrm{m}}\mathrm{Tm}$ was determined for the first time, resulting in $\mu\left({}^{152\mathrm{m}}\mathrm{Tm}\right) = 5.8(3)\,\mu_\mathrm{N}$.
Furthermore, the mean-square nuclear charge radius $\delta\langle r^2\rangle^{152\mathrm{m},169}=\SI{-1.87(17)}{\femto\meter\squared}$ \mbox{for ${}^{152\mathrm{m}}\mathrm{Tm}$} was extracted from the measured isotope shifts.}


\maketitle

\section{Introduction}\label{sec:intro}
Studies of nuclides far from the valley of $\beta$-stability advance our understanding of nuclear structure as they are particularly sensitive probes for benchmarking state-of-the-art nuclear models~\cite{Vretenar_2005, Blaum_2011, Campbell_2016}.
The chain of neutron-deficient thulium isotopes is particularly suited as it contains relatively long-lived proton emitters, e.g., ${}^{147}\mathrm{Tm}$ (\mbox{$t_{1/2}=\SI{580(30)}{\milli\second}$}~\cite{NICA20221}).
Proton emitters are nuclei with an unbound proton in the nuclear potential well that can escape via quantum tunneling, without requiring $\alpha$-particle preformation. Measuring the change in charge radius of a proton emitter therefore provides direct access to the spatial distribution of this exotic nuclear wavefunction. The thulium isotopic chain also crosses the $N=82$ shell closure at ${}^{151}\mathrm{Tm}$.
Shell closures are typically associated with spherical nuclei, thus besides neutron separation energies, the evolution of deformation parameters and charge radii is of interest to indicate the impact of shell effects. A kink in the trend of the mean-square charge radii is predicted at the $N=82$ shell closure~\cite{Perera_2021}.
Such a kink indicates a sudden change in nuclear size/shape and has already been observed in isotopic chains from tellurium to dysprosium~\cite{ANGELI201369}.
The transition region from spherical nuclei at $N=82$ to strongly deformed mid-shell nuclei also shows shape effects~\cite{Stone2021, Pramanik2016} which are influenced by the $Z=64$ subshell gap~\cite{Alkhazov1989, Garrett2022, Casten1981}.
Extending relative charge-radius measurements towards the neutron-deficient side of the thulium isotopic chain therefore provides access to the evolution of nuclear charge distribution.
In addition, the change in occupation of nuclear orbitals near the shell closure gives rise to shape coexistence and nuclear isomers which are observed in this region~\cite{Poves_2016, Nowacki_2021, Moeller_2009}.
Here, magnetic dipole moments as probe of single-particle properties are a valuable asset in evaluating the underlying nuclear structure and complementary measurements of nuclear moments provide stringent tests of density functional \mbox{theory (DFT)} calculations~\cite{Sassarini_2022,Li_2018,Reinhard_2022}.

Laser spectroscopy is particularly well suited for such studies, as it is a versatile tool for investigating nuclear properties in a nuclear-model-independent way.
Isotope-shift measurements give access to changes in the mean-square charge radii along an isotopic chain, whereas resolving the hyperfine structure (HFS) enables the determination of nuclear spins and moments from the respective hyperfine coupling constants.

Here, we report on laser spectroscopic studies of neutron deficient Tm isotopes (i.e. ${}^{152\mathrm{m}}\mathrm{Tm}$, ${}^{153}\mathrm{Tm}$, ${}^{154}\mathrm{Tm}$ and  ${}^{154\mathrm{m}}\mathrm{Tm}$) produced in fusion-evaporation reactions using the radiation detected resonance ionization spectroscopy (RADRIS) technique \cite{laatiaoui_2016, raeder_2018, warbinek_2024}. The only stable isotope, ${}^{169}\mathrm{Tm}$, has a nuclear spin of \mbox{$I=1/2$~\cite{BAGLIN20181}}, and thus any atomic transition is insensitive to the nuclear electric quadrupole deformation. Therefore, complementary off-line measurements were performed on the radionuclide ${}^{170}\mathrm{Tm}$ \mbox{($t_{1/2}=\SI{168}{\day}, I=1$)}, produced by neutron capture at the TRIGA reactor in Mainz.


\section{Off-line measurements}\label{sec:offline}
To obtain a reference of the atomic coupling parameters, laser spectroscopy was performed on ${}^{169,170}\mathrm{Tm}$ for which nuclear properties are already known. ${}^{169}\mathrm{Tm}$ was available as an inductively coupled plasma mass spectrometry  (ICP-MS) standard stock solution.
\SI{5}{\micro\L} solution containing \mbox{$10^{12}$ atoms} was deposited on a $4\times  \SI{4}{\milli\meter\squared}$ zirconium carrier foil and evaporated to dryness. The foil was then folded for insertion in an atomizer~\cite{kieck_2019}.
For the second sample, the ${}^{169}\mathrm{Tm}$ stock solution was irradiated at the research reactor TRIGA Mark II Mainz~\cite{EberhardtGeppert+2019+535+546} at the Department of Chemistry - Nuclear Chemistry at JGU Mainz, producing ${}^{170}\mathrm{Tm}$ via neutron capture reactions.
This sample, containing $2\times 10^{12}$ atoms of ${}^{169}\mathrm{Tm}$ and $8\times 10^{9}$ atoms of ${}^{170}\mathrm{Tm}$ in \SI{2}{\micro\L} solution, was also deposited on a zirconium carrier foil.

\subsection{RISIKO separator}
The off-line measurements were performed at the RISIKO mass separator~\cite{kieck_2019} at the Institute of Physics at JGU Mainz, using resonance ionization spectroscopy.
To produce neutral thulium atoms, the sample foil was folded and placed inside a tubular tantalum atomizer and resistively heated up to around \SI{2400}{\degreeCelsius}.
Laser excitation and ionization was performed in the atomic beam immediately downstream of the atomizer inside the perpendicularly illuminated-laser ion source and trap (PI-LIST)~\cite{PILIST2016, PILIST_Kron_2020}.
It consists of two repeller electrodes to suppress surface ions followed by a radio frequency quadrupole (RFQ). The PI-LIST therefore reduces ion background and Doppler broadening.
After leaving the RFQ structure, the singly-charged ions are accelerated to $\SI{30}{\kilo\eVolt}$.
Subsequently, they are separated according to their mass-to-charge ratio by passing through a \SI{60}{\degree} sector field dipole magnet, with separator slits used to suppress adjacent-mass contamination.
The typical mass resolving power of the RISIKO mass separator is \mbox{$M \slash \Delta M \geq 400$}~\cite{kieck_2019}. A secondary electron multiplier (SEM) in single ion counting mode was used to detect the ions.

The laser system consisted of two conventional ns pulsed Ti:sapphire lasers, one narrow bandwidth injection-locked Ti:sapphire laser, and a wide-range tunable Ti:sapphire laser, all homemade. Each laser was pumped by an individual Nd:YAG pump laser emitting pulses at the second harmonic at $\SI{532}{\nano\meter}$ with a repetition rate of $\SI{10}{\kilo\hertz}$. The pump lasers were  synchronized with an external pulse-pattern generator.
The injection-locked Ti:sapphire laser was used for the first excitation step (FES), seeded by an external-cavity continuous-wave (cw) diode laser (TA:pro laser; TOPTICA Photonics AG).
This injection-locked Ti:sapphire laser has a linewidth of approximately $\SI{20}{\mega\hertz}$~\cite{Sonnenschein_2017}, which is close to the Fourier limit. The fundamental output of the laser was focused into a barium borate crystal for single-pass second harmonic generation (SHG). Here, around \SI{500}{\milli\watt} of average laser power with a pulse length of \SI{50}{\nano\second} could be achieved. The output power after the SHG could be adjusted by means of a half-wave plate. Fast stabilization of the TA:pro laser was done with an iScan unit (TEM Messtechnik GmbH). To correct for a long-term frequency drift, fringe-offset locking was applied by recording the transmission of the TA:pro and a commercially available He-Ne laser (SIOS SL-03) through a home-built scanning Fabry-Pérot interferometer (S-FPI) with a finesse of $\mathcal{F}\approx 400$ and a free spectral range of \mbox{$\mathrm{FSR}=\SI{299.721}{\mega\hertz}$}. For the determination of the absolute frequency, a wavelength meter (HighFinesse WSU-30) was utilized. The absolute frequency measurement of the ionization laser was done via another wavelength meter (HighFinesse WS-600).
The broad bandwidth Ti:sapphire laser, used for the second excitation step (SES) leading to ionization, is equipped with a grating in Littrow configuration, which ensures mode-hop-free tuning over the complete Ti:sapphire range \cite{Teigelhoefer2010}. Intracavity SHG was applied, resulting in up to \SI{1.3}{\watt} average output power, a pulse length of \SIrange{40}{60}{\nano\second}, and a bandwidth of 1 to \SI{5}{\giga\hertz}.

\subsection{Off-line high-resolution spectroscopy of \texorpdfstring{${}^{169,170}\mathrm{Tm}$}{169,170Tm}}
\label{sec:high_resolution_spec}

\begin{figure}[t]
	\centering \includegraphics[width=1 \columnwidth]{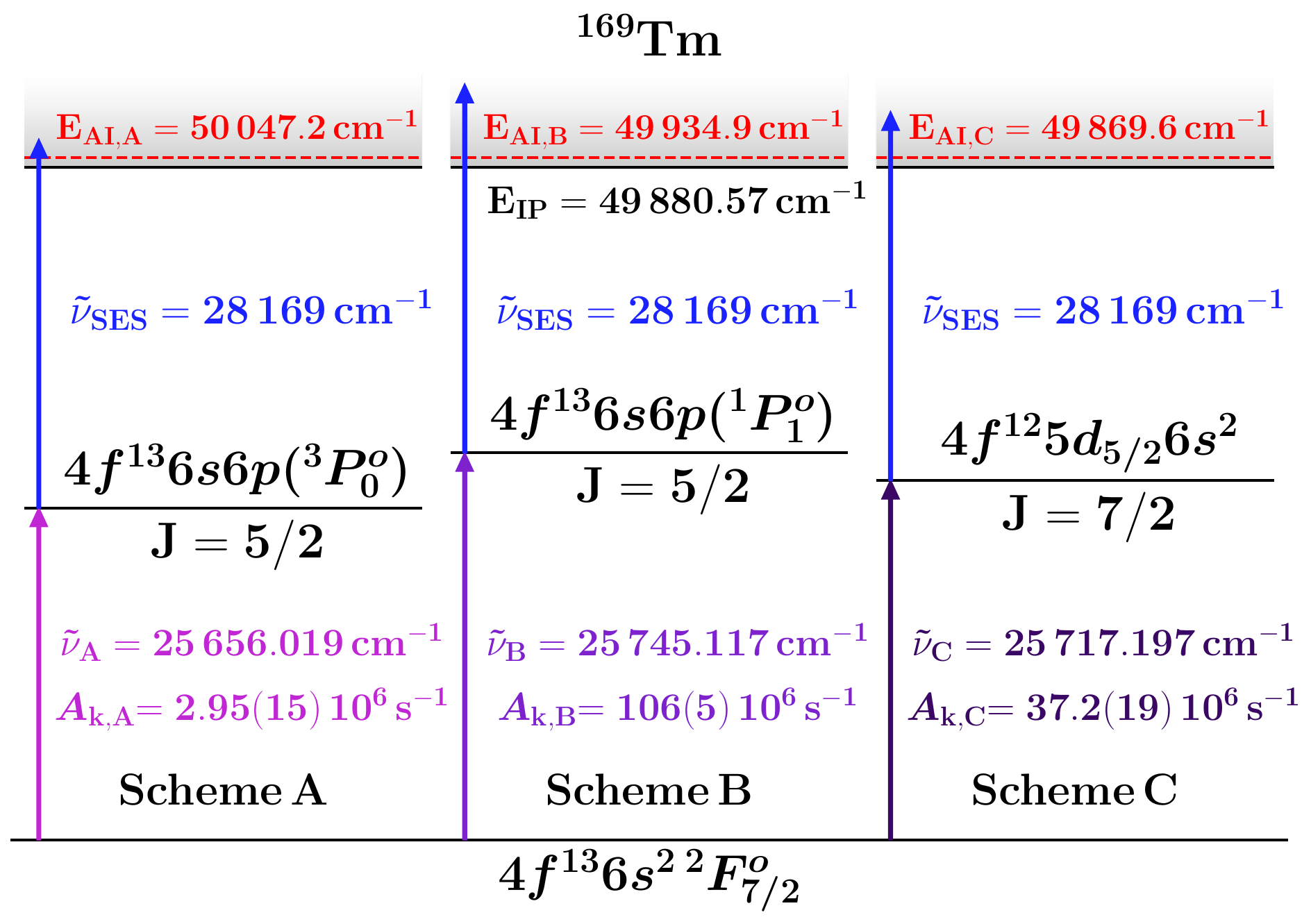}
	\caption{Two-step ionization schemes for the investigated transitions in thulium. Resonance wavenumbers $\tilde{\nu}$, their respective Einstein coefficients $A_{\mathrm{k},i}$ and configuaration assignments were taken from refs.~\cite{Sugar1973, Wickliffe1997}. The ionization potential was taken from~\cite{Mostamand_2022}.}
	\label{fig:excitation_scheme}
\end{figure}

Starting from the $4f^{13}6s^2\,{}^2F_{7/2}^{\mathrm{o}}$ ground-state configuration, two-step resonance ionization schemes were employed to investigate two ${}^2F_{7/2}^{\mathrm{o}}\rightarrow J=5/2$ transitions at $\tilde{\nu}_{\mathrm{A,FES}}=\SI{25656.019}{\per\centi\meter}$(transition A) and $\tilde{\nu}_{\mathrm{B,FES}}=\SI{25745.117}{\per\centi\meter}$ (transition B) and one ${}^2F_{7/2}^{\mathrm{o}}\rightarrow J=7/2$ transition at $\tilde{\nu}_{\mathrm{C,FES}}=\SI{25717.197}{\per\centi\meter}$ (transition C), respectively, as shown in Fig.~\ref{fig:excitation_scheme}.
Suitable resonant ionization steps were identified by initially scanning
the grating Ti:sapphire laser, thereby increasing the ionization efficiency.
Strong auto-ionizing states were identified for all
three ground-state transitions A to C as at $E_{\mathrm{AI,A}}=\SI{50047.2}{\per\centi\meter}$, \mbox{$E_{\mathrm{AI,B}}=\SI{49934.9}{\per\centi\meter}$}, and $E_{\mathrm{AI,C}}=\SI{49869.6}{\per\centi\meter}$.

To resolve hyperfine components, the experimental linewidth was minimized by optimizing the power of the spectroscopy laser and the delay between this and the ionization laser pulses. To suppress decoherence broadening caused by simultaneous excitation and ionization~\cite{de_Groote_2017}, a temporal delay between the two laser pulses of 10 to \SI{100}{\nano\second} was introduced. The values were chosen based on the excited-state lifetimes of transitions A, B, and C and the associated reduction in count rate. The corresponding lifetimes are \SI{339(17)}{\nano\second}, \SI{9.4(5)}{\nano\second}, and \SI{26.9(13)}{\nano\second}, respectively~\cite{Wickliffe1997}. The spectroscopy laser power was reduced to average powers between 5 and \SI{28}{\milli\watt} and linewidths (FWHM) of 100 to \SI{150}{\mega\hertz}, where no saturation broadening was observed. The signal remained approximately a factor of 20 above the background. The average ionization laser power was fixed at about \SI{500}{\milli\watt}. 

The HFS spectra were recorded by tuning the seed laser of the spectroscopy laser, the TA:pro laser, while monitoring the ion signal. At each frequency step, data were acquired for 1 to \SI{2}{\second}, depending on the counting statistics. To minimize wavelength drifts, the wavelength meter was calibrated with the Rb-locked external-cavity diode laser (ECDL) before and after each scan. In addition, each HFS spectrum was recorded in both scanning directions to account for possible scan-direction-dependent frequency drifts of the Ti:sapphire lasers.

\begin{figure}[t]
	\centering \includegraphics[width=1 \columnwidth]{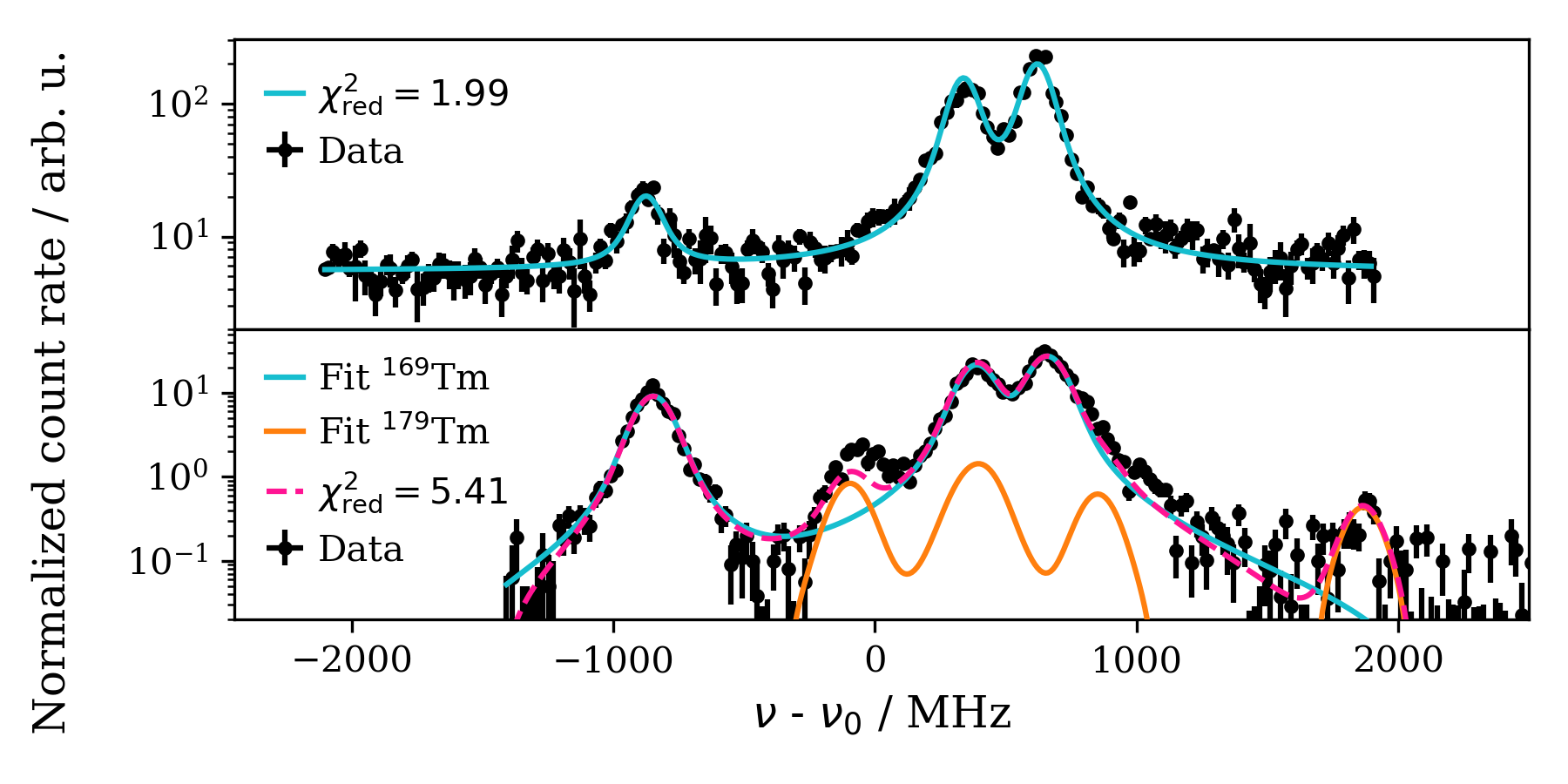}
	\caption{Measured hyperfine structure for transition A for ${}^{169}\mathrm{Tm}$ (top) as well as ${}^{169}\mathrm{Tm}$ and ${}^{170}\mathrm{Tm}$ (bottom). The centroid frequency of ${}^{169}\mathrm{Tm}$ stated in Fig.~\ref{fig:excitation_scheme} is taken as reference for the frequency offset $\nu-\nu_0$. Fit parameters are given in Tab.~\ref{tab:hfs_parameters}.}
	\label{fig:hfs_transition_A}
\end{figure}
The recorded spectra for transition A are shown in Fig.~\ref{fig:hfs_transition_A}. The spectrum of ${}^{170}\mathrm{Tm}$ contains contributions from ${}^{169}\mathrm{Tm}$, as the abundance of ${}^{169}\mathrm{Tm}$ in the ion source exceeded that of ${}^{170}\mathrm{Tm}$ by approximately 2.5 orders of magnitude. The neighbouring isotope ${}^{169}\mathrm{Tm}$ could not be completely suppressed by the mass separator, resulting in a contaminant signal approximately one order of magnitude stronger than the ${}^{170}\mathrm{Tm}$ signal. Nevertheless, the achieved linewidth of approximately \SI{100}{\mega\hertz} was sufficient to resolve hyperfine components.
The HFS spectra were fitted with the \texttt{satlas} Python package~\cite{Satlas}, which allows fitting hyperfine spectra using a sum of Voigt profiles. The angular momenta of the transitions and nuclear spins were fixed to their literature values. To estimate the statistical uncertainty, the recorded datasets were fitted using two different binning sizes. Owing to the higher counting statistics, the uncontaminated ${}^{169}\mathrm{Tm}$ spectrum was fitted first to determine the lower- and upper-state hyperfine constants $\mathcal{A}_l$ and $\mathcal{A}_u$, respectively, which were then used to constrain the fit of the composite spectrum of ${}^{170}\mathrm{Tm}$ and ${}^{169}\mathrm{Tm}$.
The scans were performed with a laser power of \SI{5}{\milli\watt} and \SI{28}{\milli\watt}, and delays of the second step of 10 and \SI{100}{\nano\second}, respectively.
For the result of the HFS constants, the weighted average of the resulting fit parameters was taken with the standard deviation used as the uncertainty.
The determined HFS constants $\mathcal{A}_\mathrm{l}^{169}$ and $\mathcal{A}_\mathrm{u}^{169}$ as well as the literature values are listed in Tab.~\ref{tab:hfs_parameters}.
Our results for both constants agree with the literature values.
Since the HFS spectra of the individual isotopes were recorded simultaneously under identical experimental conditions, the spectrum containing both ${}^{169}\mathrm{Tm}$ and ${}^{170}\mathrm{Tm}$ was fitted with the sum of two HFS models, sharing common FWHM and saturation parameters.
Furthermore, the HFS constants $\mathcal{A}$ of the upper and lower states of ${}^{169}\mathrm{Tm}$ were fixed to the fitted value from the HFS spectrum of the isotopically pure ${}^{169}\mathrm{Tm}$.
The HFS constant $\mathcal{B}_\mathrm{l}$ of the lower state of ${}^{170}\mathrm{Tm}$ was fixed to the literature value (see Tab.~\ref{tab:hfs_parameters}).
The HFS constants $\mathcal{A}_{\mathrm{u}}^{170}$ or $\mathcal{A}_{\mathrm{l}}^{170}$ were fixed using the relation
\begin{align} \label{eq:mu_from_reference}
    \mu &= \frac{\mathcal{A} \cdot I}{\mathcal{A}_{\text{ref}} \cdot I_{\text{ref}}} \cdot \mu_{\text{ref}},
\end{align}
where $\mathcal{A}_{\mathrm{ref}}$ is the HFS constant of the lower or upper state of ${}^{169}\mathrm{Tm}$, respectively.
The magnetic moments were taken from literature and are listed in Tab.~\ref{tab:nuclear_dipole_moments}.
For the final value of $\mathcal{B}_{\mathrm{u}}^{\mathrm{170}}$, the weighted mean of the fit results was calculated.
The quoted uncertainty corresponds to the standard deviation.
The results are listed in Tab.~\ref{tab:hfs_parameters}.
An additional systematic uncertainty of \SI{8}{\mega\hertz} was added to the hyperfine constants $\mathcal{A}$ and $\mathcal{B}$.
It reflects the maximum observed deviation of the wavelength meter readout from the FPI during its characterization and was conservatively \mbox{adopted ~\cite{Verlinde2020}}.
With the determined HFS constants, the ratios $\mathcal{A}_{\mathrm{l}}^{\mathrm{170}}/\mathcal{A}_{\mathrm{u}}^{\mathrm{170}}$ and $\mathcal{B}_{\mathrm{l}}^{\mathrm{170}}/\mathcal{B}_{\mathrm{u}}^{\mathrm{170}}$ can be calculated.
For each individual transition these ratios should stay constant over the isotopic chain and thus can be used to constrain fit parameters assuming hyperfine anomalies are negligible within our experimental uncertainties.
For transition A, the ratios were determined to \mbox{$\mathcal{A}_{\mathrm{l}}^{\mathrm{170}}/\mathcal{A}_{\mathrm{u}}^{\mathrm{170}} = 0.915(31)$} and \mbox{$\mathcal{B}_{\mathrm{l}}^{\mathrm{170}}/\mathcal{B}_{\mathrm{u}}^{\mathrm{170}} = 3.247(115)$}.
For the other transitions, ${}^{170}\mathrm{Tm}$ could not be observed.
Therefore, no $\mathcal{B}$-ratio for transitions B and C could be determined.
The centroid positions $\nu_\mathrm{A}^{170}$ and $\nu_\mathrm{A}^{169}$, as well as $\nu_\mathrm{B}^{169}$ and $\nu_\mathrm{C}^{169}$, relative to the literature value of ${}^{169}\mathrm{Tm}$, are summarized in Tab.~\ref{tab:hfs_parameters}. 
$\nu_\mathrm{A}^{169}$ was obtained by taking the weighted averages of the fit parameters of the scans with the unirradiated and irradiated sample.
For $\nu_\mathrm{B}^{169}$ and $\nu_\mathrm{C}^{169}$ only the scans with the unirradiated sample were taken into account.
The uncertainties of the centroid positions listed in Tab.~\ref{tab:hfs_parameters} correspond to one standard deviation of the individual measurements.
Additional systematic uncertainties were added: \SI{60}{\mega\hertz}, specified by the manufacturer of the wavelength meter and \SI{25}{\mega\hertz} to account for the Doppler shift caused by a possible deviation from a perpendicular alignment between the laser and atomic beams. 
The measured centroid position of ${}^{169}\mathrm{Tm}$ deviates from the literature reference by $\SI{510(70)}{\mega\hertz}$ for transition A. The uncertainties reported for the three transitions in ref.~\cite{Sugar1973} range from \SIrange{120}{300}{\mega\hertz}. Thus, our result and the literature values agree within the $1.5\sigma$ to $3\sigma$ uncertainty range.
Since the observed offset is reproducible across different scans of the same transition but differs between transitions (see Tab.~\ref{tab:hfs_parameters}), a systematic frequency shift of our setup is unlikely to account for the observed discrepancy. Furthermore, an independent measurement performed to determine the pressure shift (see App.~\ref{Appendix:pressure_shift}) yielded a centroid position of $\SI{350(60)}{\mega\hertz}$ for transition A. This value agrees with our result within $1.5\sigma$.
For transition B, our result agrees with the literature value within the $1.5\sigma$ to $2.5\sigma$ uncertainty range and is also consistent with the centroid position of $\SI{350(60)}{\mega\hertz}$ obtained in the pressure-shift measurement. For transition C, our measured centroid position agrees within uncertainty with both the literature value and the result of the pressure-shift measurement, $\SI{30(60)}{\mega\hertz}$ (see App.~\ref{Appendix:pressure_shift}).
The consistent agreement between our measurements and the independent pressure-shift measurements for all three transitions, together with the transition-dependent deviations from the literature values, suggests that the discrepancy is not caused by a systematic shift in our experiment. In particular, the agreement observed for transition C provides no indication of a systematic offset. Therefore, the literature centroid position for transition A, and potentially also for transition B, may require reconsideration.

\section{On-line measurements}\label{sec:Online}
The nuclides ${}^{152\mathrm{m}}\mathrm{Tm}$, ${}^{153}\mathrm{Tm}$, and ${}^{154\mathrm{m}}\mathrm{Tm}$ were produced in complete fusion-evaporation reactions of a $^{52}\mathrm{Cr}$ beam on a $^{107}\mathrm{Ag}$ target.
The beam was provided by the UNILAC accelerator at GSI Helmholtzzentrum für Schwerionenforschung, Darmstadt, at an energy of \SI{4.35}{\mega\eVolt} per nucleon with an average current of $I_\mathrm{B} \approx 6 \times 10^{11}$ particles per second, delivered at a repetition rate of \SI{5}{\hertz} from the \SI{50}{\hertz} accelerator cycle.
The areal density of the silver targets was \SI{0.4}{\milli\gram\per\centi\meter\squared}.
The evaporation residues were separated in-flight from the primary beam by the velocity filter SHIP~\cite{MUNZENBERG197965, block2022recent}.

\subsection{RADRIS setup}
After separation, the ions entered the buffer-gas stopping cell RADRIS, located downstream of the SHIP focal plane.
They passed through \SI{8}{\micro\meter} of Mylar degrader foils and a \SI{3.5}{\micro\meter}-thick aluminium-coated Mylar entrance window before entering the gas cell.
The slowed-down ions were fully thermalized in ultrahigh-purity argon buffer gas at \SI{95}{\milli\bar}. For neutralization, they were accumulated on a \SI{25}{\micro\meter}-thick, \SI{1}{\milli\meter}-wide hafnium strip filament that was biased to an attractive potential~\cite{raeder_2018, chhetri_2018, warbinek_2022}.
The duration of this accumulation phase was chosen based on the half-life of the ion of interest~\cite{laatiaoui_2014, warbinek_2022, laatiaoui_2013} and was set to \SI{8}{\second}. The beam was subsequently switched off and the filament was resistively heated for \SI{300}{\milli\second} to 1350--1400~$^\circ$C.
Upon release, an atom cloud forms around the filament and is ionized in a two-step laser resonant ionization scheme.
The first excitation step was scanned with a tunable dye laser (Liop-Tec LiopStar-HQ, bandwidth \SI{\sim 4}{\giga\hertz}, pulse width \SI{10}{\nano\second}) pumped by the third harmonic of a Nd:YAG laser.
The laser beam was transported to the experimental setup via optical fibres.
The laser light for the second excitation step was provided by a second Nd:YAG laser (Innolas Spitlight DPSS) with an average energy per pulse of approximately \SI{70}{\milli\joule} and a pulse width of approximately \SI{10}{\nano\second}.
Both Nd:YAG lasers were operated at a repetition rate of \SI{100}{\hertz} and were synchronized with an external pulse generator.
Their beam profiles were spatially superimposed, each having a diameter of approximately 2 to \SI{3}{\centi\meter}.
Laser frequencies were monitored using a wavelength meter (HighFinesse WS7), which was calibrated at least once per day with a frequency-stabilized He-Ne laser.
Since ${}^{152\mathrm{m}}\mathrm{Tm}$ does not exhibit a usable $\alpha$-decay branch, it was identified indirectly after its $\beta^+$ decay ($t_{1/2}=\SI{5.2(6)}{\second}$) into ${}^{152}\mathrm{Er}$. 
The subsequent $\alpha$ decay of ${}^{152}\mathrm{Er}$ was detected with a silicon detector.
The erbium decays with a half-life of \SI{10.3(1)}{\second} predominantly by $\alpha$ emission (branching ratio $\mathrm{b}_\alpha=\SI{90(4)}{\percent}$) to ${}^{148}\mathrm{Dy}$ with an $\alpha$-particle energy of \SI{4804.3(16)}{\kilo\eVolt}~\cite{Toth_1987}.

\subsection{Identification of populated nuclear states}\label{sec:states}

\begin{figure}[tb]
	\centering \includegraphics[width=\columnwidth]{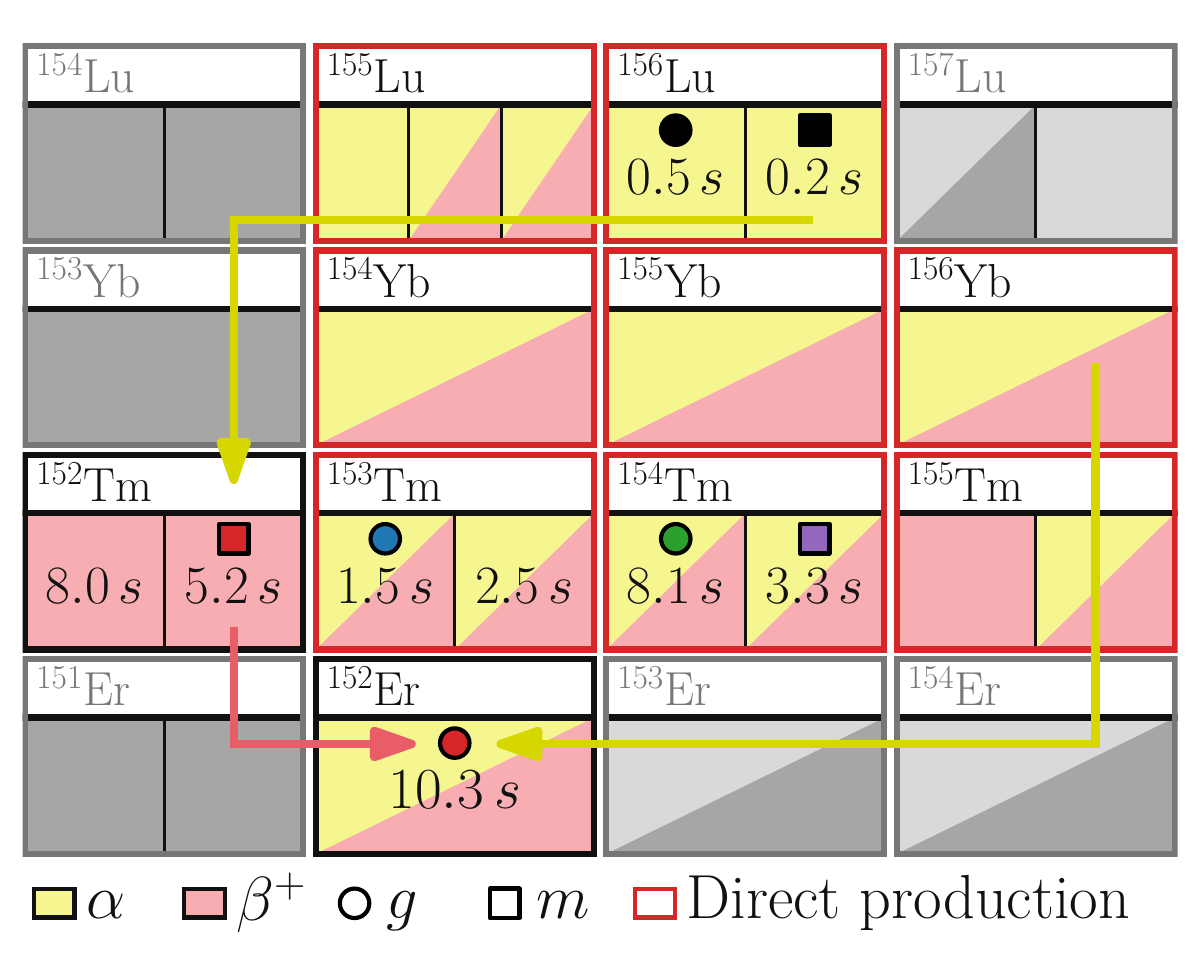}
	\caption{Excerpt of the nuclear chart showing isotopes relevant to production, laser spectroscopy, and detection.
	Decay modes are color-coded, and half-lives are indicated for ground and isomeric states.
	Colored circles and squares mark ground and isomeric states of interest, respectively.
	Red frames denote directly produced isotopes.
	Annotated arrows show decay paths.}
	\label{fig:nuclear_chart_excerpt}
\end{figure}

\begin{figure}[tb]
	\centering \includegraphics[width=\columnwidth]{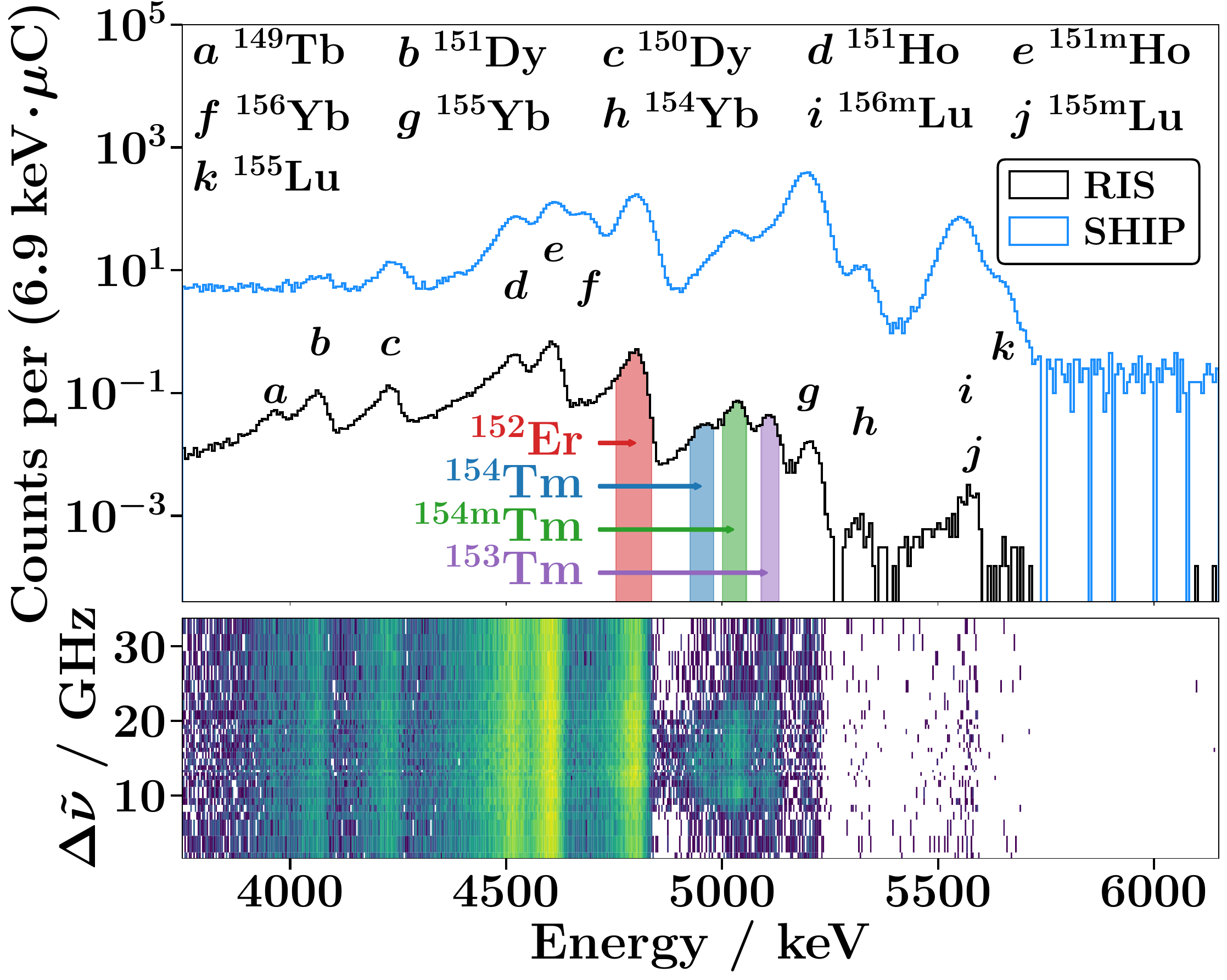}
	\caption{\textbf{Top}: $\alpha$ spectrum measured at the focal plane of SHIP (blue) and cumulative $\alpha$ spectrum (black) acquired as transition A was scanned. Colored regions indicate the energy gates used for the laser spectra shown in Fig.~\ref{fig:spectrum_25717_25745}.
	The count rate was normalized to the primary-beam dose. In total, $9.4(8)\times10^{12}$ and $3.1(2)\times10^{15}$ ${}^{52}\mathrm{Cr}$ ions were delivered during the SHIP and RIS measurements, respectively.
	\textbf{Bottom}: Heat map of the energy-binned $\alpha$-decay counts as a function of energy and frequency offset $\Delta\nu=\nu-\nu_0$, highlighting spectral features that vary with the laser frequency.}
	\label{fig:25656_combined}
\end{figure}

To investigate whether isomeric states in ${}^{152-154}\mathrm{Tm}$ have been populated by the $^{107}\mathrm{Ag}(^{52}\mathrm{Cr}, x\mathrm{n})$ fusion-evaporation reaction, the $\alpha$-energy spectrum acquired while scanning transition A (see Fig.~\ref{fig:25656_combined}) and results from heavy-ion vaporation (HIVAP)~\cite{Sagaidak_2013} calculations (cf. Fig.~\ref{fig:hivap}) were analyzed. An overview of nuclei relevant for production, laser spectroscopy, and detection can be found in Fig.~\ref{fig:nuclear_chart_excerpt}.
The isotope ${}^{152}\mathrm{Tm}$ is found to be predominantly in the isomeric state ${}^{152\mathrm{m}}\mathrm{Tm}$, as it is produced via the $\alpha$ decay of ${}^{156\mathrm{m}}\mathrm{Lu}$. The \SI{5565}{\kilo\eVolt} $\alpha$-decay line of ${}^{156\mathrm{m}}\mathrm{Lu}$, which populates the isomer, is clearly observed in the spectrum, whereas ${}^{156\mathrm{g}}\mathrm{Lu}$, which would feed ${}^{152\mathrm{g}}\mathrm{Tm}$, is not observed.
Direct production and feeding of ${}^{152}\mathrm{Tm}$ via ${}^{152}\mathrm{Yb}$ are not expected based on isotopic distributions and HIVAP calculations.
However, ${}^{152}\mathrm{Er}$, which was used to identify ${}^{152}\mathrm{Tm}$, is fed by the $\mathrm{b}_\alpha=\SI{10}{\percent}$ $\alpha$-branch of ${}^{156}\mathrm{Yb}$ ($t_{1/2}=\SI{26.1(7)}{\second}$), leading to a constant background on this $\alpha$ energy.
For ${}^{154(\mathrm{m})}\mathrm{Tm}$, possible indirect feeding through the $\beta$ decay of ${}^{154}\mathrm{Yb}$ or the $\alpha$ decay of ${}^{158\mathrm{m}}\mathrm{Lu}$ is expected to be negligible.
Thus ${}^{153}\mathrm{Tm}$ and ${}^{154}\mathrm{Tm}$ are produced directly in the $\alpha x\mathrm{n}$-evaporation channels.
For ${}^{154}\mathrm{Tm}$, the isomeric and the ground state can be discriminated by their characteristic $\alpha$-decay energies.
Here, predominantly the high-spin isomer $(I^\pi = 9^+)$ is produced rather than the low-spin ground state $(I^\pi = 2^+)$.
For ${}^{153}\mathrm{Tm}$, the ground state and isomeric state have similar half-lives and $\alpha$-decay energies, but they differ significantly in their spin, with $I^{\pi}=(11/2)^-$ for the ground state and $I^{\pi}=(1/2)^+$ for the isomeric state.
Analogous to ${}^{156}\mathrm{Lu}$ and ${}^{154}\mathrm{Tm}$, the main population is therefore expected in the high-spin ground state.
This assumption is in agreement with the results of the HFS analysis as discussed in the following.

\subsection{Broadband spectroscopy of the neutron-deficient thulium isotopes}\label{sec:broadband_spec}
\begin{figure}[b]
	\centering \includegraphics[width=1 \columnwidth]{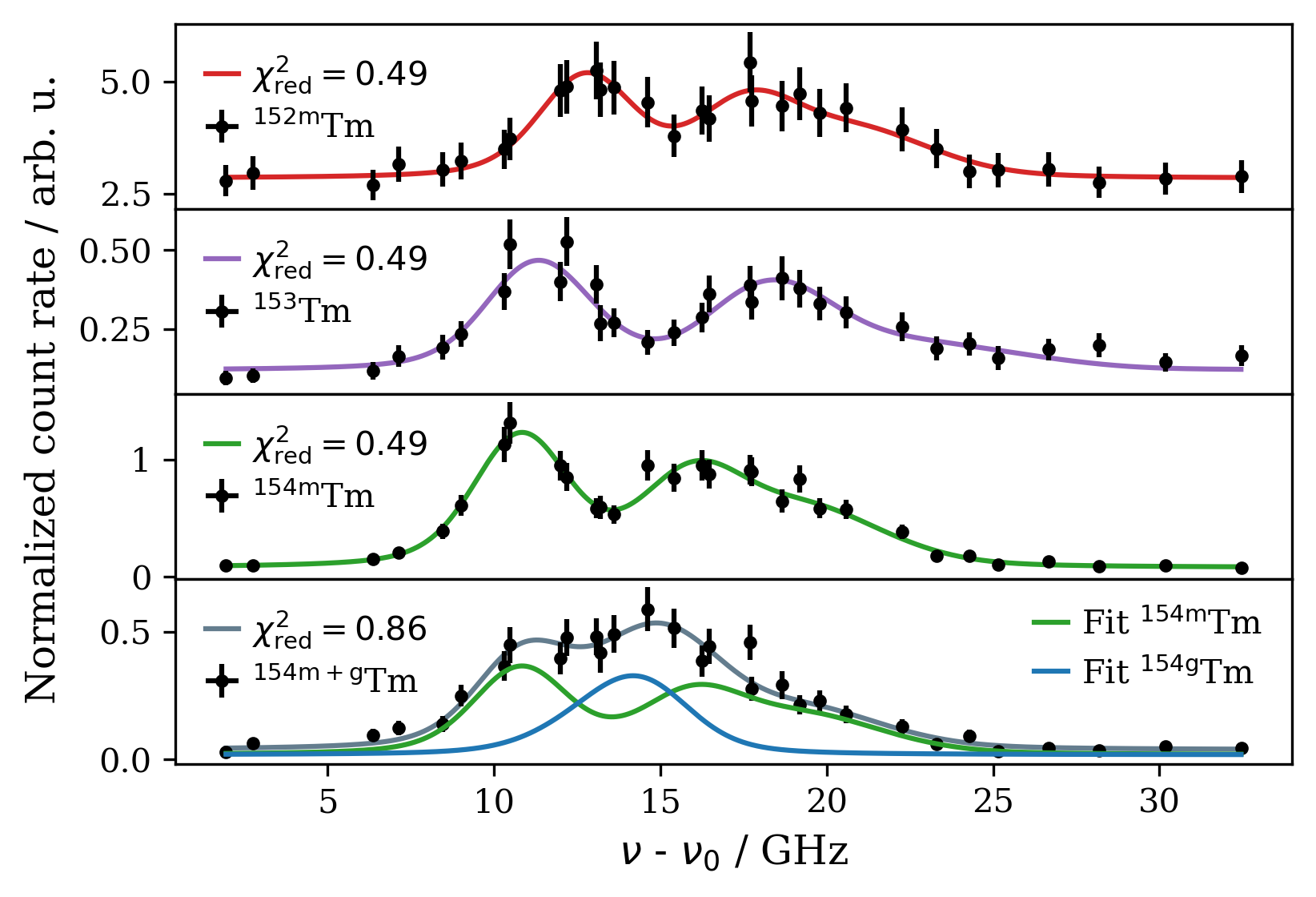}
	\caption{Measured hyperfine structure for transition A for ${}^{152\mathrm{m}}\mathrm{Tm}$, ${}^{153}\mathrm{Tm}$, and ${}^{154\mathrm{m}}\mathrm{Tm}$ as well as measured spectrum for ${}^{154\mathrm{m}}\mathrm{Tm}$ and ${}^{154\mathrm{g}}\mathrm{Tm}$ (bottom). The centroid frequency of ${}^{169}\mathrm{Tm}$ stated in Fig.~\ref{fig:excitation_scheme} is taken as reference for the frequency offset $\nu-\nu_0$. Fit parameters are given \mbox{in Tab.~\ref{tab:hfs_parameters}}.}
	\label{fig:spectrum_25656}
\end{figure}

\begin{figure}[t]
	\centering \includegraphics[width=1 \columnwidth]{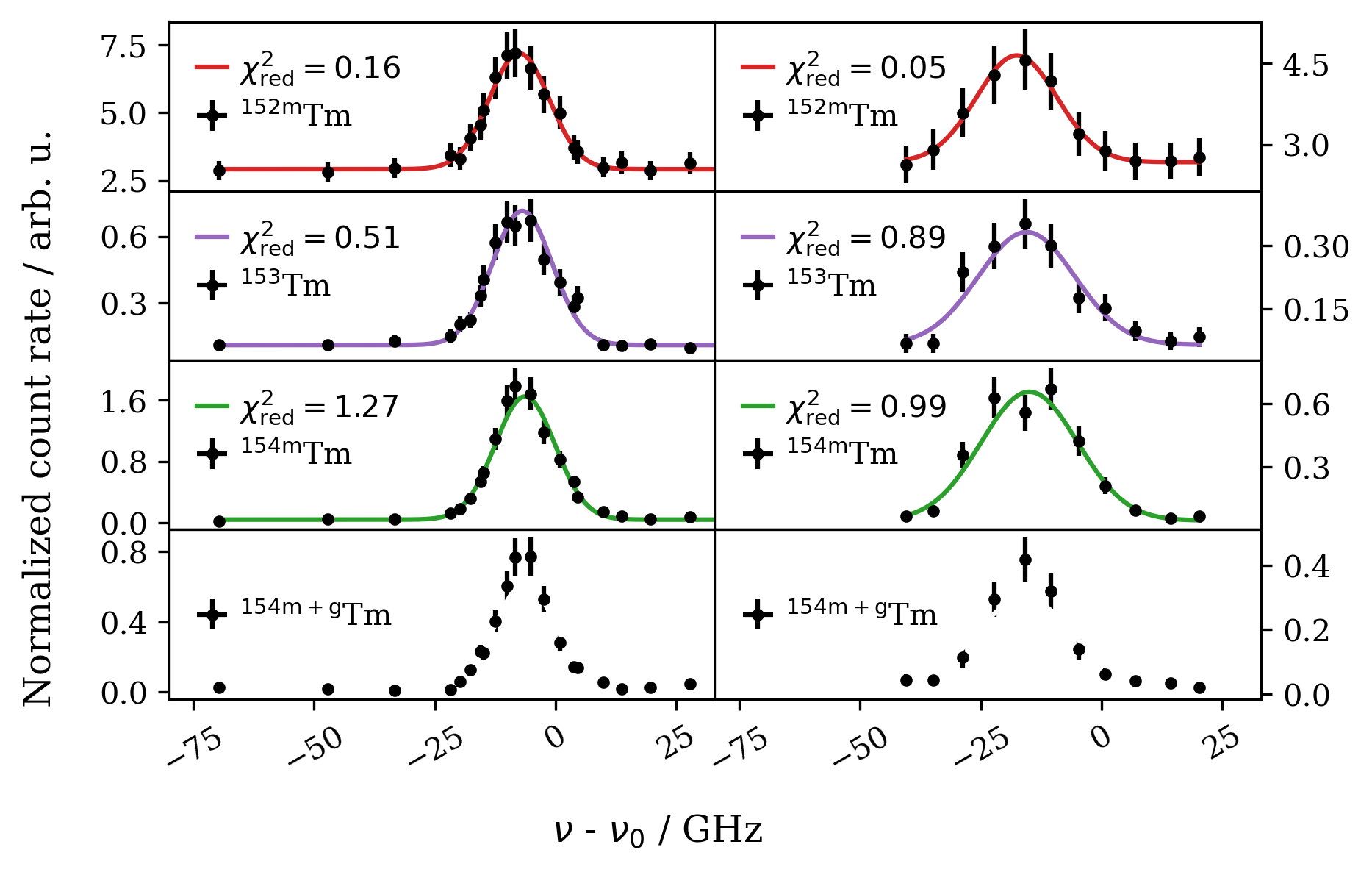}
	\caption{ Measured spectra for transition B (left) and transition C (right) for ${}^{152\mathrm{m}}\mathrm{Tm}$, ${}^{153}\mathrm{Tm}$, and ${}^{154\mathrm{m}}\mathrm{Tm}$. The centroid frequency of ${}^{169}\mathrm{Tm}$ stated in Fig.~\ref{fig:excitation_scheme} is taken as reference for the frequency offset $\nu-\nu_0$. Fit parameters are given \mbox{in Tab.~\ref{tab:hfs_parameters}.}}
	\label{fig:spectrum_25717_25745}
\end{figure}

For the on-line measurements of the neutron-deficient thulium isotopes, namely ${}^{152\mathrm{m}}\mathrm{Tm}$, ${}^{153}\mathrm{Tm}$, ${}^{154}\mathrm{Tm}$, and ${}^{154\mathrm{m}}\mathrm{Tm}$, the same optical transitions as in Sec.~\ref{sec:offline}, i.e. transitions A, B, and C, were studied. 
The transitions were first scanned with an average power of around \SI{10}{\milli\watt}
(\SI{100}{\micro\joule} per pulse) to locate the resonance position.
Due to the linewidth of up to \SI{5}{\giga\hertz}, the HFS was not resolved for these scans.
The later scans of transition A, however, were then performed with a lower average laser power of \SI{5}{\milli\watt} to reduce power broadening effects. 
For the second, ionizing step, the high power Nd:YAG laser at \SI{355}{\nano\meter} was used to ionize the atoms non-resonantly with a pulse energy of roughly \SI{70}{\milli\joule}.
To account for the broadening mechanism described in Sec.~\ref{sec:high_resolution_spec}, the delay was set to \SI{10}{\nano\second} relative to the first step.
To accumulate sufficient statistics and average over multiple heating pulses, the measurement cycle was repeated 40 times for each laser frequency step.
Before proceeding with the next laser frequency, a waiting time of \SI{40}{s}, corresponding approximately to five half-lives of the longest-lived investigated nuclide, was used to allow the radionuclides on the detector to decay and thereby reduce unwanted background. 
Gating on the $\alpha$ spectra as discussed before allowed to extract the optical spectra of transition A, which can be seen in Fig.~\ref{fig:spectrum_25656}.
The scans of transitions B and C are shown in Fig.~\ref{fig:spectrum_25717_25745}.
At least five scans of transition A and one scan of transition B and C were recorded where all isotopes were measured simultaneously.
The similar $\alpha$-decay energies of ${}^{154\mathrm{m}}\mathrm{Tm}$ and ${}^{154}\mathrm{Tm}$ result in a contribution from ${}^{154\mathrm{m}}\mathrm{Tm}$ to the spectrum assigned to ${}^{154}\mathrm{Tm}$.
For all scans, the spectral resolution was limited primarily by the resolution of about \SI{2}{\giga\hertz} of the scanning laser, collisional, Doppler, and power broadening, leading to a FWHM of \SI{5}{\giga\hertz} for the high power scans and \SI{3}{\giga\hertz} for the lower power scans. For transition A, the HFS structure was partially resolved showing a double-peak structure as observed in Fig.~\ref{fig:spectrum_25656}.
The \texttt{satlas} package~\cite{Satlas} was used to fit the spectra.
Despite the presence of 18 allowed transitions for the spin $I=11/2$ and spin $I=9$ isotopes, only a double-peak structure was resolved. Therefore, the number of free fit parameters had to be reduced by imposing constraints.
For this, the previously determined HFS constant ratios, \mbox{$\mathcal{A}_{\mathrm{l}}^{\mathrm{170}}/\mathcal{A}_{\mathrm{u}}^{\mathrm{170}} = 0.915(31)$} and \mbox{$\mathcal{B}_{\mathrm{l}}^{\mathrm{170}}/\mathcal{B}_{\mathrm{u}}^{\mathrm{170}} = 3.247(115)$}, were used.
Since all the neutron-deficient isotopes were measured simultaneously, the FWHM and saturation parameters were shared among the fits of the HFS spectra of the individual isotopes.
During the fitting procedure, the total angular momentum of the atomic ground and excited state, as well as the nuclear spins were fixed to literature values (cf. Tab.~\ref{tab:nuclear_dipole_moments}). Since the ${}^{154}\mathrm{Tm}$ spectrum contains contributions from ${}^{154\mathrm{m}}\mathrm{Tm}$, it was fitted using a sum of HFS models. The parameters of ${}^{154\mathrm{m}}\mathrm{Tm}$ were fixed to the values obtained from the fit of the pure ${}^{154\mathrm{m}}\mathrm{Tm}$ spectrum, except for the amplitude. The HFS constants of the lower state were fixed to the literature values found in ref.~\cite{Seliverstov_2000}, while those of the upper state were constrained using the ratios determined previously. Consequently, for ${}^{154}\mathrm{Tm}$, only the amplitude and the centroid position remained free fit parameters.
\begin{table*}[t]
\caption{Extracted parameters from the hyperfine spectra of transition A. The subscripts $\mathrm{l}$ and $\mathrm{u}$ denote the associated lower and upper level. Centroid positions $\nu_i$ are stated relative to the values in Fig.~\ref{fig:excitation_scheme}. For the hyperfine constants $\mathcal{A}$ and $\mathcal{B}$, an additional systematic uncertainty of \SI{8}{\mega\hertz}, originating from the wavelength meter calibration~\cite{Verlinde2020}, has been added. For the centroid positions of the off-line data, two additional systematic uncertainties of \SI{60}{\mega\hertz} and \SI{25}{\mega\hertz} apply. For the centroid positions of the on-line data, an additional systematic uncertainty of \SI{150}{\mega\hertz} applies. For details on the fixed and dependent parameters, see text.}
\label{tab:hfs_parameters}
\begin{threeparttable}
\begin{tabular}{
    c
    @{\hspace{3mm}}
    c
    @{\hspace{3mm}}
    c
    @{\hspace{3mm}}
    c
    @{\hspace{3mm}}
    c
    @{\hspace{3mm}}
    c
    @{\hspace{3mm}}
    c
    @{\hspace{3mm}}
    c
    @{\hspace{3mm}}
    }
 \hline
 Isotope  &  $\mathcal{A}_\mathrm{A,l}$&  $\mathcal{A}_\mathrm{A,u}$&   $\mathcal{B}_\mathrm{A,l}$&   $\mathcal{B}_\mathrm{A,u}$ &  $\nu_\mathrm{A}$ &  $\nu_\mathrm{B}$ &   $\nu_\mathrm{C}$ \\
  & [MHz] & [MHz]& [MHz]& [MHz]& [MHz]& [MHz]& [MHz]  \\
  \hline\hline
  \multicolumn{8}{l}{This work} \\ \hline
  ${}^{152\mathrm{m}}\mathrm{Tm}$  &  520(27) & 569$^b$ & 150(1650)  & 50$^b$ & 16300(200) & -7540(270) & -17600(400)  \\
  ${}^{153}\mathrm{Tm}$ &  1003(72) & 1096$^b$ & -1470(3830)  & -450$^b$ & 15200(300) & -6910(310) & -15500(1100)\\
  ${}^{154\mathrm{m}}\mathrm{Tm}$ &  548(29) & 599$^b$ & 550(1140) & 170$^b$ & 14500(200) & -6320(290) & -14900(700)\\
  ${}^{154}\mathrm{Tm}$ &  -460$^c$ & -503$^b$ & -600$^c$ & -185$^b$ & 14200(200) & \textemdash & \textemdash \\
  ${}^{169}\mathrm{Tm}$  & -377(8)$^a$  & -412(11) & 0  & 0 & 510(70) & 460(70) & 90(70)\\
  ${}^{170}\mathrm{Tm}$ & 201$^b$ & 219$^b$ & -1006.7$^c$ & -310(12) & 430(70) & \textemdash & \textemdash \\\hline
  \multicolumn{4}{l}{Literature} &  References\\ \hline
  ${}^{153}\mathrm{Tm}$ & 1020(16) & & -700(1300) & \cite{Seliverstov_2000} & & & \\
  ${}^{154\mathrm{m}}\mathrm{Tm}$  & 532(3) &  & 200(500) & \cite{Seliverstov_2000} & & & \\
  ${}^{154}\mathrm{Tm}$  & -460(10) &  & -600(1200) & \cite{Seliverstov_2000} & & & \\
  ${}^{169}\mathrm{Tm}$ & -374.1374(16) & -410.7(14) & 0 & \cite{Ritter_1962, Pfeufer_1986} & & &   \\
  ${}^{170}\mathrm{Tm}$ & 199.0(6) &  & -1006.7(18)  & \cite{Dyer_1988} & &\\ 
  \hline
\end{tabular}
\begin{tablenotes}
\small
\item[a] Determined only using scheme A.
\item[b] Dependent parameter: Determined with Eq.~(\ref{eq:mu_from_reference}) by taking the fit value of ${}^{169}\mathrm{Tm}$ as the reference or fixing HFS constant ratios.
\item[c] Value fixed to literature value. 
\end{tablenotes}
\end{threeparttable}
\end{table*}
For the other spectra exhibiting only a single peak due to broadening, the line shape was fitted with a Voigt profile and a constant background.
The spectrum of $^{152\mathrm{m}}\mathrm{Tm}$ was analyzed first as it exhibited the highest counting statistics.
The FWHM values of the peaks extracted from the fits were subsequently fixed for the analysis of the other isotopes, since the isotopes were recorded simultaneously and are therefore subject to the same broadening mechanisms. For ${}^{154}\mathrm{Tm}$, no centroid positions could be obtained due to the contribution of ${}^{154\mathrm{m}}\mathrm{Tm}$ in the spectrum.
The fit results, including the extracted HFS constants and centroid positions, for both the single-peak spectra and the partially resolved HFS spectra are summarized in Tab.~\ref{tab:hfs_parameters}. The final values were obtained as the weighted mean of the individual fit results, with the uncertainties given by the corresponding standard deviations and adding systematic uncertainties as discussed in Sec.~\ref{sec:high_resolution_spec}. For the centroid positions, a systematic uncertainty of \SI{150}{\mega\hertz} is considered here instead of the \SI{60}{\mega\hertz} used in Sec.~\ref{sec:high_resolution_spec}, as multi-mode fibers were used in these measurements.
For two of the scans with lower laser powers, no HFS constants and centroid positions could be extracted, as the fits did not converge.
The centroid position for transition A was determined by only considering the double-peak spectra.
Furthermore, as the on-line measurements were performed in a gaseous environment, the pressure shift resulting from collisional interactions between the laser-excited states and the Ar buffer gas at 95 mbar had to be accounted for.
A dedicated measurement yielded a pressure-shift coefficient for transition A of $\gamma_\mathrm{sh}=\SI{-2.0(4)}{\mega\hertz\per\milli\bar}$ \cite{Harsh_2026}, corresponding to a (red) shift of $\Gamma_\mathrm{sh}=\SI{-190(40)}{\mega\hertz}$ of the on-line resonance centroids (see App.~\ref{Appendix:pressure_shift}).

\section{Discussion}\label{sec:discussion}
\subsection{Nuclear moments}
The two most precise values for nuclear dipole moments in $\mathrm{Tm}$ exist for ${}^{169}\mathrm{Tm}$ and ${}^{170}\mathrm{Tm}$ and are listed in Tab.~\ref{tab:nuclear_dipole_moments}.
These were measured with different methods.
The value for ${}^{169}\mathrm{Tm}$ was measured with the atomic beam magnetic resonance method~\cite{Giglberger1967} and was later corrected with a new calculated diamagnetic correction \cite{BAGLIN20082033}.
For ${}^{170}\mathrm{Tm}$, the value stated in reference~\cite{BAGLIN20181} is a weighted mean measured in experiments using the technique of $\beta$-radiation-detected optical pumping and atomic beam resonance fluorescence.
The magnetic moments of the investigated neutron-deficient isotopes, ${}^{152\mathrm{m}}\mathrm{Tm}$, ${}^{153}\mathrm{Tm}$, and ${}^{154\mathrm{m}}\mathrm{Tm}$, can be determined with the measured $\mathcal{A}$ parameter using Eq.~(\ref{eq:mu_from_reference}) with ${}^{169}\mathrm{Tm}$ as reference isotope.
The results are listed in Tab.~\ref{tab:nuclear_dipole_moments}. 

\begin{table}[ht]
\centering
\caption{Magnetic dipole moments of nuclei used in this work.}
\label{tab:nuclear_dipole_moments}
\begin{tabular}{lcccc}
\hline
Isotope & $I^\pi$ & $\mu$ & $\mu_\mathrm{Lit}$ & Refe- \\
& & [$\mu_\mathrm{N}$] & [$\mu_\mathrm{N}$] & rence\\
\hline
& & This work & Literature & \\
${}^{152\mathrm{m}}\mathrm{Tm}$ & $9^+$ & $+5.8(3)$ & & \\
${}^{153}\mathrm{Tm}$ & $(11/2^-)$ & $+6.8(5)$ & $+6.93(11)$ & \cite{Seliverstov_2000} \\
${}^{154\mathrm{m}}\mathrm{Tm}$ & $9^+$ & $+6.1(2)$ & $+5.91(5)$ & \cite{Seliverstov_2000} \\
${}^{154}\mathrm{Tm}$ & $2^-$ & & $-1.14(2)$ & \cite{Seliverstov_2000} \\
${}^{169}\mathrm{Tm}$ & $1/2^+$ & & $-0.2316(15)$ & \cite{BAGLIN20082033} \\
${}^{170}\mathrm{Tm}$ & $1^-$ & & $+0.2468(12)$ & \cite{BAGLIN20181} \\
\hline
${}^{145}\mathrm{Eu}$ & $11/2^-$ & & $+7.46(4)$ & \cite{Stone_2011} \\
${}^{147}\mathrm{Eu}$ & $11/2^-$ & & $+7.05(3)$ & \cite{Stone_2011} \\
${}^{149}\mathrm{Eu}$ & $11/2^-$ & & $+7.0(3)$ & \cite{Stone_2011} \\
${}^{153}\mathrm{Er}$ & $7/2^-$ & & $-0.934(7)$ & \cite{Stone_2011} \\
${}^{151}\mathrm{Dy}$ & $7/2^-$ & & $-0.945(7)$ & \cite{Stone_2011} \\
${}^{155}\mathrm{Yb}$ & $7/2^-$ & & $-0.91(2)$ & \cite{Stone_2011} \\
${}^{145}\mathrm{Gd}$ & $11/2^-$ & & $-1.0(2)$ & \cite{Stone_2011} \\
${}^{147}\mathrm{Gd}$ & $7/2^-$ & & $-1.02(9)$ & \cite{Stone_2011} \\
${}^{149}\mathrm{Gd}$ & $7/2^-$ & & $-0.88(4)^{\textrm{a}}$ & \cite{Stone_2011} \\
${}^{147}\mathrm{Sm}$ & $7/2^-$ & & $-0.8148(7)$ & \cite{Stone_2011} \\
\hline
\end{tabular}
\begin{tablenotes}
\small
\item[a] Negative sign was inferred as the original work only determined $\vert \mu \vert$ \cite{Kracikova_1987}.
\end{tablenotes}
\end{table}

The obtained magnetic dipole moments of ${}^{153}\mathrm{Tm}$ and ${}^{154\mathrm{m}}\mathrm{Tm}$ are in good agreement with the literature. It should be noted that, recalculating the literature magnetic moments of ${}^{153}\mathrm{Tm}$ and ${}^{154\mathrm{m}}\mathrm{Tm}$ using the hyperfine constants $\mathcal{A}_{\mathrm{l}}^{\mathrm{153}}$ and $\mathcal{A}_{\mathrm{l}}^{\mathrm{154m}}$ reported in the original work~\cite{Seliverstov_2000}, together with the literature values of $\mathcal{A}_{\mathrm{l}}^{\mathrm{169}}$ and $\mu\left({}^{169}\mathrm{Tm}\right)$~\cite{BAGLIN20082033}, yields slightly altered values of \mbox{$\mu\left({}^{153}\mathrm{Tm}\right)=+6.95(12)\,\mu_N$} and \mbox{$\mu\left({}^{154\mathrm{m}}\mathrm{Tm}\right)=+5.93(5)\,\mu_N$}.
A small contribution from the hyperfine anomaly ${}^i\Delta^j$, which describes deviations from Eq.~(\ref{eq:mu_from_reference}), i.e. \mbox{$\mathcal{A}_j/\mathcal{A}_i = g^i _I / g^j_I (1-{}^i\Delta^j)$} where $g_I = \mu/I$ is the gyromagnetic ratio for the respective isotopes $i$ and $j$, cannot be excluded. However, hyperfine anomalies are typically below \SI{1}{\percent}~\cite{PERSSON2023101589} and are therefore expected to be covered by the quoted uncertainties.
The good agreement obtained for ${}^{153}\mathrm{Tm}$ and ${}^{154\mathrm{m}}\mathrm{Tm}$ indicates that no significant systematic bias is present and supports the reliability of the magnetic moment determined for ${}^{152\mathrm{m}}\mathrm{Tm}$.

The obtained magnetic dipole moment of ${}^{152\mathrm{m}}\mathrm{Tm}$ ($I=9$) is very close to the values reported for ${}^{154\mathrm{m}}\mathrm{Tm}$ ($I=9$), ${}^{152\mathrm{m}}\mathrm{Ho}$ ($I=9$, $\mu=5.94(5)\,\mu_N$), and ${}^{148\mathrm{m}}\mathrm{Eu}$ \mbox{($I=9$, $\mu=6.120(45)\,\mu_N$)}, indicating that the same orbitals may be involved in the formation of these states. A natural assumption is the proton $\pi h_{11/2}$ orbital, which is progressively filled in the proton-rich rare-earth region, coupled to the $\nu f_{7/2}$ neutron orbital. The $g$ factors of the $\pi h_{11/2}$ proton orbital are $g({}^{153}\mathrm{Tm})=1.260(20)$, \mbox{$g({}^{147}\mathrm{Eu})=1.282(5)$}, $g({}^{149}\mathrm{Eu})=1.27(5)$, and $g({}^{145}\mathrm{Eu})=1.356(7)$, while the corresponding $g$ factors of the $\nu f_{7/2}$ neutron orbital are $g({}^{153}\mathrm{Er})=-0.267(2)$, \mbox{$g({}^{151}\mathrm{Dy})=-0.270(2)$}, $g({}^{155}\mathrm{Yb})=-0.260(6)$, $g({}^{147}\mathrm{Gd})=-0.291(26)$, $g({}^{149}\mathrm{Gd})=-0.251(11)$, $g({}^{147}\mathrm{Sm})=-0.2328(2)$, and $g({}^{145}\mathrm{Gd})=-0.29(6)$ (cf. Tab.~\ref{tab:nuclear_dipole_moments}). Applying the additivity rule for fully aligned spins and using the average of these values gives a $g$-factor of $g(9^+)=0.67$, corresponding to $\mu(9^+)=6.05\,\mu_N$, in excellent agreement with the observed magnetic moments of these states.

\subsection{Isotope shifts/nuclear mean-square charge radius}
The isotope shift (IS) is defined as the difference in the transition frequency between two isotopes with mass $A$ and $A'$,
\begin{equation}
    \delta \nu_{i}^{A,A'} = \nu^{A'}_i - \nu^{A}_i,
\end{equation}
where $i$ denotes the investigated atomic transition and $A$ indicates a fixed reference isotope. 
When calculating the isotope shifts from the centroid positions listed in Tab.~\ref{tab:hfs_parameters}, we obtain $\delta\nu_{\mathrm{A}}^{A,A'}<0$, $\delta\nu_{\mathrm{B}}^{A,A'}>0$, and $\delta\nu_{\mathrm{C}}^{A,A'}>0$ for $A>A'$. The sign of the isotope shift in general is sensitive to the change of electronic configurations. Transitions promoting an electron with angular momentum $l$ to an orbital with $l'>l$ are generally expected to exhibit $\delta\nu_i^{A,A'}<0$ for $A>A'$ due to the decreased electron density at the nucleus for the excited state. For a change $l'<l$, analogously, a shift of $\delta\nu_i^{A,A'}>0$ for $A>A'$ is expected. The observed signs for transitions A and C are consistent with these expectations. In contrast, transition B exhibits a positive isotope shift, whereas the literature assignment~\cite{Sugar_1973} would suggest a negative value. This discrepancy indicates that the upper level of transition B may have been assigned incorrectly in the literature or it is heavily mixed. 

To determine the changes in mean-square charge radii from the data, the isotope shift can be expressed as
\begin{align}
	\delta \nu_{i}^{A,A'} = M^{-1}\cdot\,K_i + F_i \cdot \delta \langle r^2\rangle^{A, A'},
	\label{eq:isotope_shift}
\end{align}
with $M^{-1} = (m^{A'}-m^{A}) / (m^{A'} \cdot m^{A})$.
The isotope shift is composed of two contributions: the \mbox{mass shift $M^{-1}K_i$}, originating from the finite nuclear mass, and the \mbox{field shift $F_i\cdot \delta \langle r^2\rangle^{A, A'}$}, reflecting the finite size and shape of the nucleus.
The field-shift factor $F_i$ and mass-shift factor $K_i$ are transition-dependent atomic quantities that can be obtained from atomic calculations~\cite{Cheal2012} or determined experimentally using a modified King plot when charge radii are already known in the literature.
Multiplying Eq.~(\ref{eq:isotope_shift}) by $M$ yields a linear relation between the modified isotope shift $M\delta\nu_{i}^{A,A'}$ and the modified change in mean-square charge radius $M\delta\langle r^2\rangle^{A,A'}$, called King plot \cite{King2013}, with $F_i$ and $K_i$ corresponding to the slope and intercept, respectively.
Thus, isotope shifts of isotopes with known $\delta\langle r^2\rangle^{A,A'}$ can be used to determine the atomic factors and subsequently determine unknown $\delta\langle r^2\rangle^{A,A'}$ for the remaining isotopes.
To calculate the relative mean-square charge radius differences, the literature values of \mbox{$\delta \langle r^2\rangle^{170,169}=\SI{0.048(1)}{\femto\meter\squared}$}, $\delta \langle r^2\rangle^{154,169}=\SI{-1.609(15)}{\femto\meter\squared}$, and \mbox{$\delta \langle r^2\rangle^{153,169}=\SI{-1.708(33)}{\femto\meter\squared}$} from ref.~\cite{ANGELI201369} were used to determine the field and mass shift factors from a King plot analysis.
\begin{figure}[b]
	\centering \includegraphics[width=1 \columnwidth]{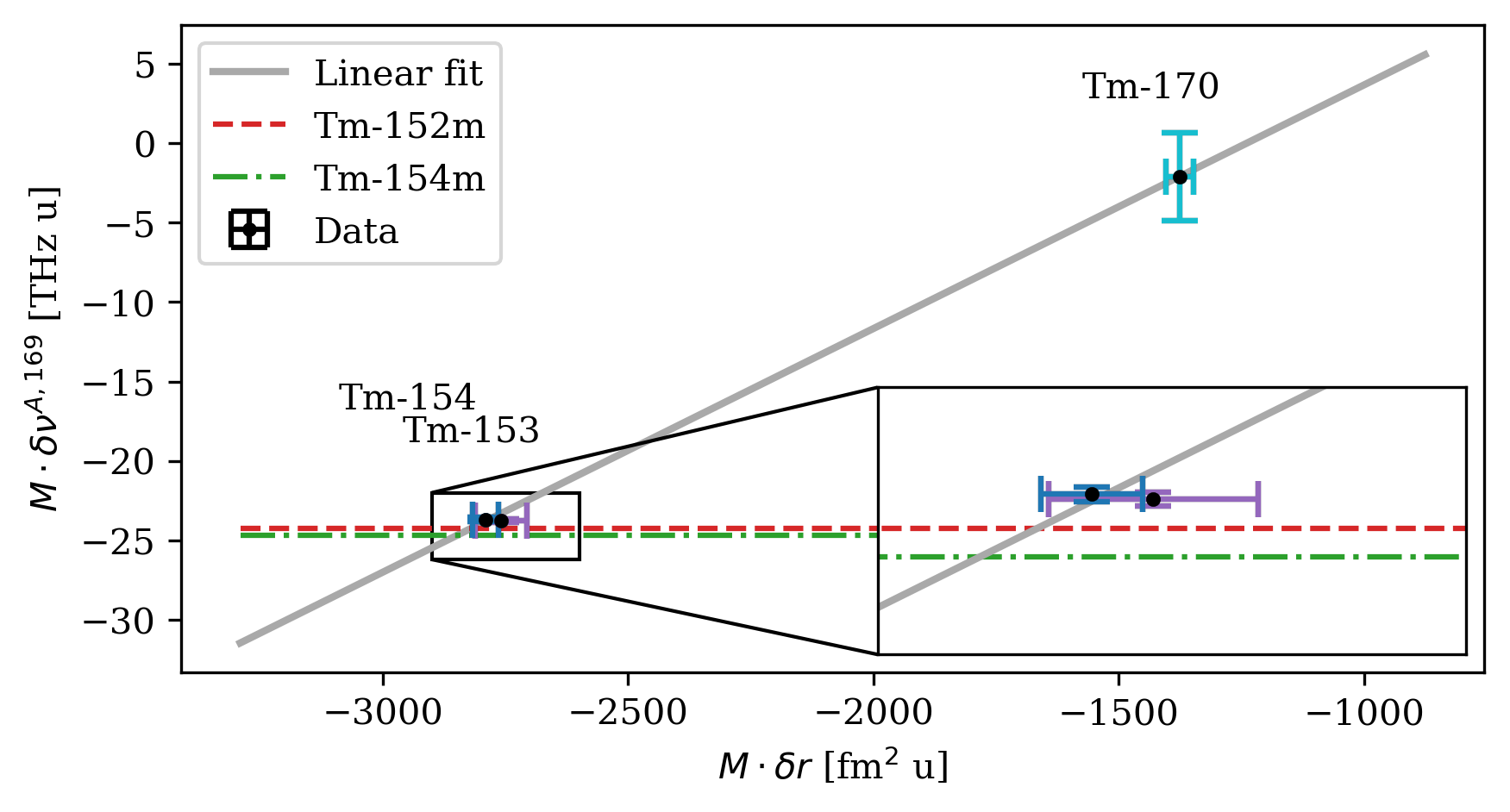}
	\caption{King plot for transition A showing modified isotope shifts as a function of the modified charge radii differences. The horizontal lines indicate the modified isotope shift from the measured centroid positions of the isotopes for which the change in mean-square charge radius has been determined.}
	\label{fig:kingplot}
\end{figure}
The isotope shifts were calculated using the centroid positions shown in Tab.~\ref{tab:hfs_parameters}. 
The corrected centroid for the pressure shift was used to calculate the modified isotope shifts shown in the King plot \mbox{in Fig.~\ref{fig:kingplot}}.
Using orthogonal distance regression, a field shift factor of \mbox{$F_\mathrm{A} = \SI{15.3(11)}{\giga\hertz\per\femto\meter\squared}$} and a mass shift factor of \mbox{$K_\mathrm{A} = \SI{19.0(3.4)}{\tera\hertz\cdot u}$} are determined for scheme A. The uncertainties are given by the fit uncertainties.
Transitions B and C were not considered for the determination of the mean-square charge radius differences of ${}^{152\mathrm{m}}\mathrm{Tm}$ and ${}^{154\mathrm{m}}\mathrm{Tm}$, as the centroid positions of ${}^{170}\mathrm{Tm}$ could not be measured for these transitions.
Finally, using Eq.~(\ref{eq:isotope_shift}) yields 

\begin{align*}
    \delta \langle r^2\rangle^{152\mathrm{m}, 169} &= \SI{-1.87(17)}{\femto\meter\squared}\quad\mathrm{and}\\
    \delta \langle r^2\rangle^{154\mathrm{m}, 169} &= \SI{-1.64(15)}{\femto\meter\squared}.
\end{align*}

\begin{figure}[t]
	\centering \includegraphics[width=\columnwidth]{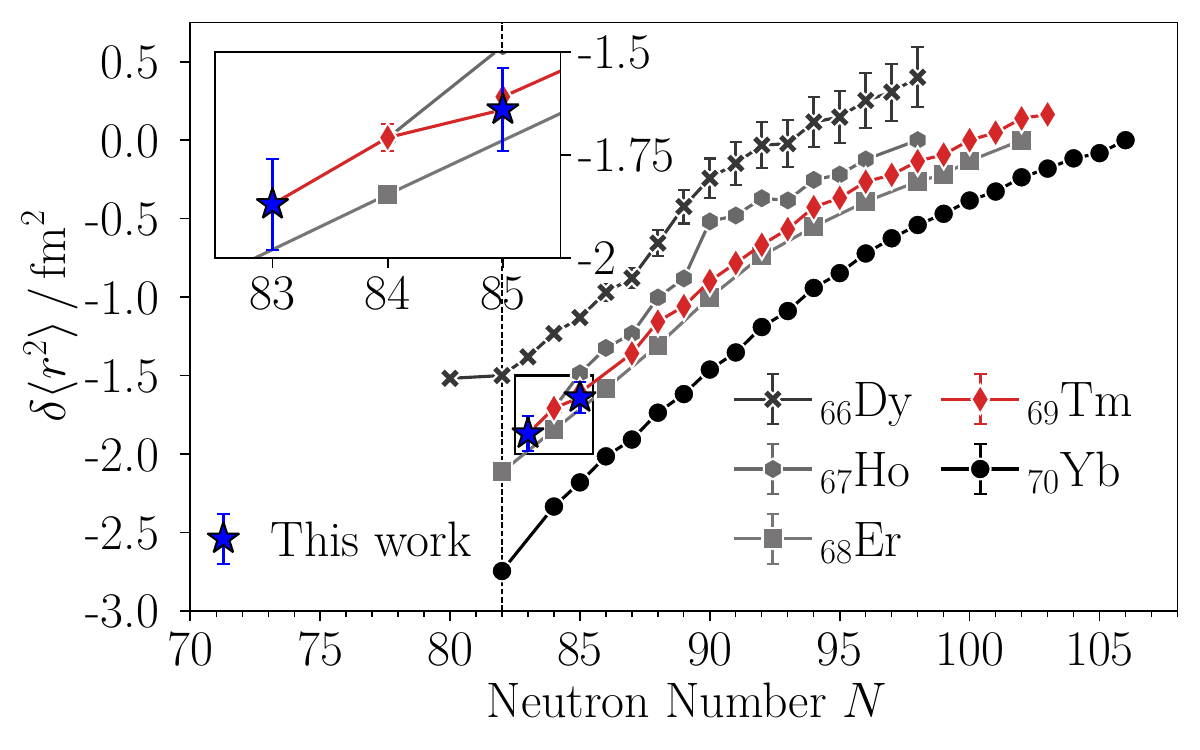}
	\caption{Changes in mean-square charge radii in the thulium region.
	Data for the neighbouring elements are taken from~\cite{ANGELI201369}. The dashed line indicates the $N=82$ shell closure. For better visibility, an offset of \SI{-1.5}{\femto\meter\squared} was added to the dysprosium (Dy) data.}
	\label{fig:charge_radii}
\end{figure}

As expected, the mean-square charge radii shown in Fig.~\ref{fig:charge_radii} decrease with decreasing neutron number. Systematic studies of the neighbouring rare-earth elements Er, Yb, Ho, and Dy have established that the nuclei evolve from nearly spherical shapes at the $N=82$ shell closure towards well-deformed prolate ground states with increasing neutron number~\cite{Sprouse1990,Neugart1985}. Correspondingly, the charge radii exhibit a smooth increase above the shell closure, reflecting both the increasing nuclear size and the gradual onset of quadrupole deformation. In particular, the pronounced kink observed at $N=82$ in the Dy isotopic chain is attributed to the neutron shell closure and the associated reduction of the ground-state deformation.
For $\mathrm{Tm}$, the behaviour when crossing \mbox{the $N=82$} shell remains an open question, as the most neutron-deficient isotope investigated is ${}^{152\mathrm{m}}\mathrm{Tm}$ \mbox{with $N=83$}. Nevertheless, the measured charge radii follow a smooth trend. This behaviour is consistent with a gradual reduction of the ground-state deformation towards the shell closure and provides no indication of an abrupt structural change before reaching $N=82$.

\section{Conclusion}\label{sec:conclusion}
We have measured the hyperfine structure of the atomic ground-state transition at \SI{25656.019}{\per\centi\meter} for the four radioactive isotopes ${}^{152\mathrm{m}}\mathrm{Tm}$, ${}^{153}\mathrm{Tm}$, ${}^{154\mathrm{m}}\mathrm{Tm}$, and ${}^{170}\mathrm{Tm}$ as well as for the stable thulium isotope ${}^{169}\mathrm{Tm}$. For ${}^{154}\mathrm{Tm}$, the centroid position was determined.
Isotope shifts for two more atomic ground-state transitions were measured for ${}^{152\mathrm{m}}\mathrm{Tm}$, ${}^{153}\mathrm{Tm}$, ${}^{154\mathrm{m}}\mathrm{Tm}$, and ${}^{169}\mathrm{Tm}$. The magnetic moments of ${}^{153}\mathrm{Tm}$ and ${}^{154\mathrm{m}}\mathrm{Tm}$ were confirmed and the magnetic moment of ${}^{152\mathrm{m}}\mathrm{Tm}$ was determined for the first time.
The known mean-square charge radii of ${}^{153}\mathrm{Tm}$, ${}^{154}\mathrm{Tm}$, ${}^{170}\mathrm{Tm}$, and ${}^{169}\mathrm{Tm}$ were used together with the centroid frequencies (corrected for pressure shift when necessary) measured for the \SI{25656.019}{\per\centi\meter} atomic ground-state transition in a King-plot analysis to extract the change in mean-square charge radii of ${}^{154\mathrm{m}}$Tm and ${}^{152\mathrm{m}}\mathrm{Tm}$.
For ${}^{152\mathrm{m}}\mathrm{Tm}$, these measurements represent the first experimental determination of its magnetic moment and mean-square charge radius.

To improve our understanding of the shell-gap evolution \mbox{at $N=82$}, it is important to extend the laser spectroscopic studies of thulium to lighter isotopes.
Extending the studies down to ${}^{147}\mathrm{Tm}$ would be the first laser spectroscopic study of a proton-emitting nucleus and would provide access to its charge radius for the first time.

The present work demonstrates the applicability of the RIS technique in a gas cell for determining nuclear moments and charge radii of short-lived isotopes.
It provides the foundation for future studies employing the in-gas-Jet Resonance Ionization Spectroscopy (JetRIS) setup, where RIS is performed in a supersonic gas jet with a spectral resolution down to \SI{250}{\mega\hertz}~\cite{Lantis_2024, Claessens_2025}. This will then enable the extraction of the spectroscopic quadrupole deformations from the hyperfine $\mathcal{B}$ constants which were inaccessible in the present work due to the limited resolution.
Together with the ongoing implementation of a cooler-buncher and a multireflection time-of-flight mass spectrometer, JetRIS will extend these studies beyond $\alpha$-decaying nuclei by enabling mass-selective ion detection~\cite{Muenzberg2025}.
These developments will substantially broaden the range of nuclides accessible to laser spectroscopy at SHIP.

\backmatter
\bmhead{Supplementary information}
\bmhead{Acknowledgements}
We gratefully acknowledge the help of the Staff of the TRIGA reactor in Mainz for irradiating our sample. We also thank Dennis Renisch and Yuki Ishikawa for the chemical preparation of the irradiated sample. Additional recognition goes to the Target Labor of GSI, for preparing the silver Target that was used during the on-line experiment.
The results presented here are based on the experiment G-22-00051, which were performed at the beam line/target station Y7/SHIP at the GSI Helmholtzzentrum für Schwerionenforschung, Darmstadt (Germany) in the frame of FAIR \mbox{Phase-0}.
This project has received funding from the European Union’s Horizon Europe Research and Innovation programme under Grant Agreement No 101057511 (EURO-LABS).
This work has been supported by the Bundesministerium für Forschung, Technologie und Raumfahrt (BMFTR, Germany) under project numbers 05P21RDFN1 and by the GSI open access fund. DR acknowledges support from Grant No. PID2022-141496NB-I00, funded by MICIU/AEI/10.13039/501100011033 and by ERDF/EU. H, AC, RF, FI and PVD acknowledge the funding from the Research Foundation Flanders (FWO, Belgium) BOF KU Leuven (C14/22/104), and the FWO and F.R.S.-FNRS under the Excellence of Science (EOS) program (40007501).


\subsubsection*{Data Availability Statement}
Associated data can be found on Zenodo~\cite{weyrich_2026_23187436}.

\subsubsection*{Code Availability Statement}
Code/software will be made available on reasonable request.

\section*{Declarations}
The authors declare that they have no known competing financial
interests or personal relationships that could have appeared to influence
the work reported in this paper. 

\bibliography{tm}
\newpage
\begin{appendices}

\section{HIVAP calculations}\label{Appendix:HIVAP}
The production cross sections calculated with HIVAP~\cite{Sagaidak_2013} for the ${}^{52}\mathrm{Cr}+{}^{107}\mathrm{Ag}$ fusion-evaporation reaction are shown in Fig.~\ref{fig:hivap}. 
The estimated yields were used to identify the populated nuclear states as discussed in Sec.~\ref{sec:states}.
\begin{center}
	\centering \includegraphics[width= \columnwidth]{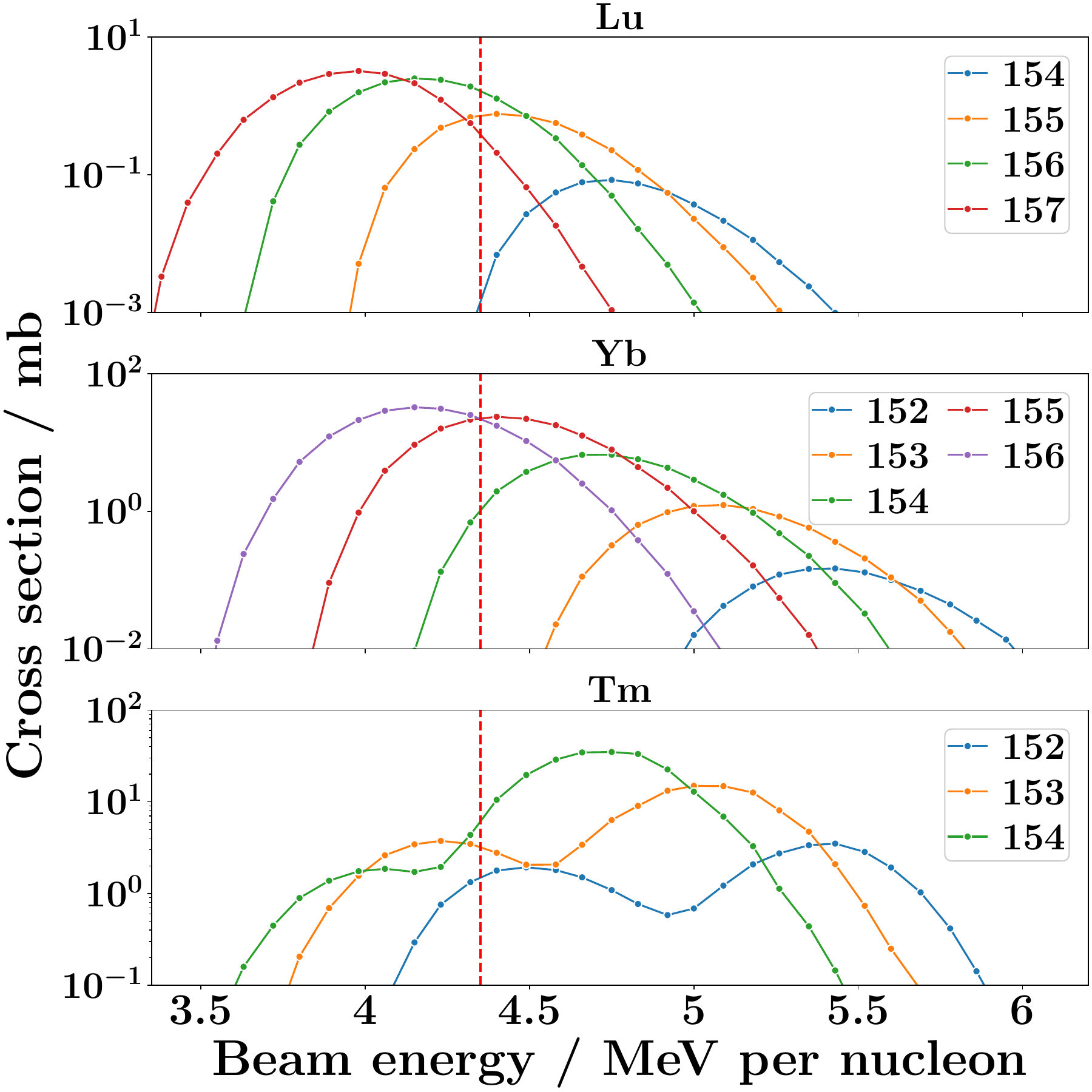}
	\captionof{figure}{\textbf{Top}: predicted cross sections of compound nuclei ${}^{154-157}\mathrm{Lu}$.
	\textbf{Middle}: predicted cross sections of ${}^{152-156}\mathrm{Yb}$, produced via the p$x$n evaporation channels.
	\textbf{Bottom}: predicted cross sections of ${}^{152-154}\mathrm{Tm}$ created through the $\alpha x\mathrm{n}$ (low-energy) and 2p$x$n (high-energy) evaporation channels.
	The dashed line at $\SI{4.35}{\mega\eVolt}$ per nucleon marks the beam energy that was used in this experiment.}
	\label{fig:hivap}
\end{center}

\newpage

\section{HFS scans of transition B and C of \texorpdfstring{${}^{169}\mathrm{Tm}$}{169Tm}}\label{Appendix:HFS Scan}

\begin{figure}[htb]
	\centering \includegraphics[width= \columnwidth]{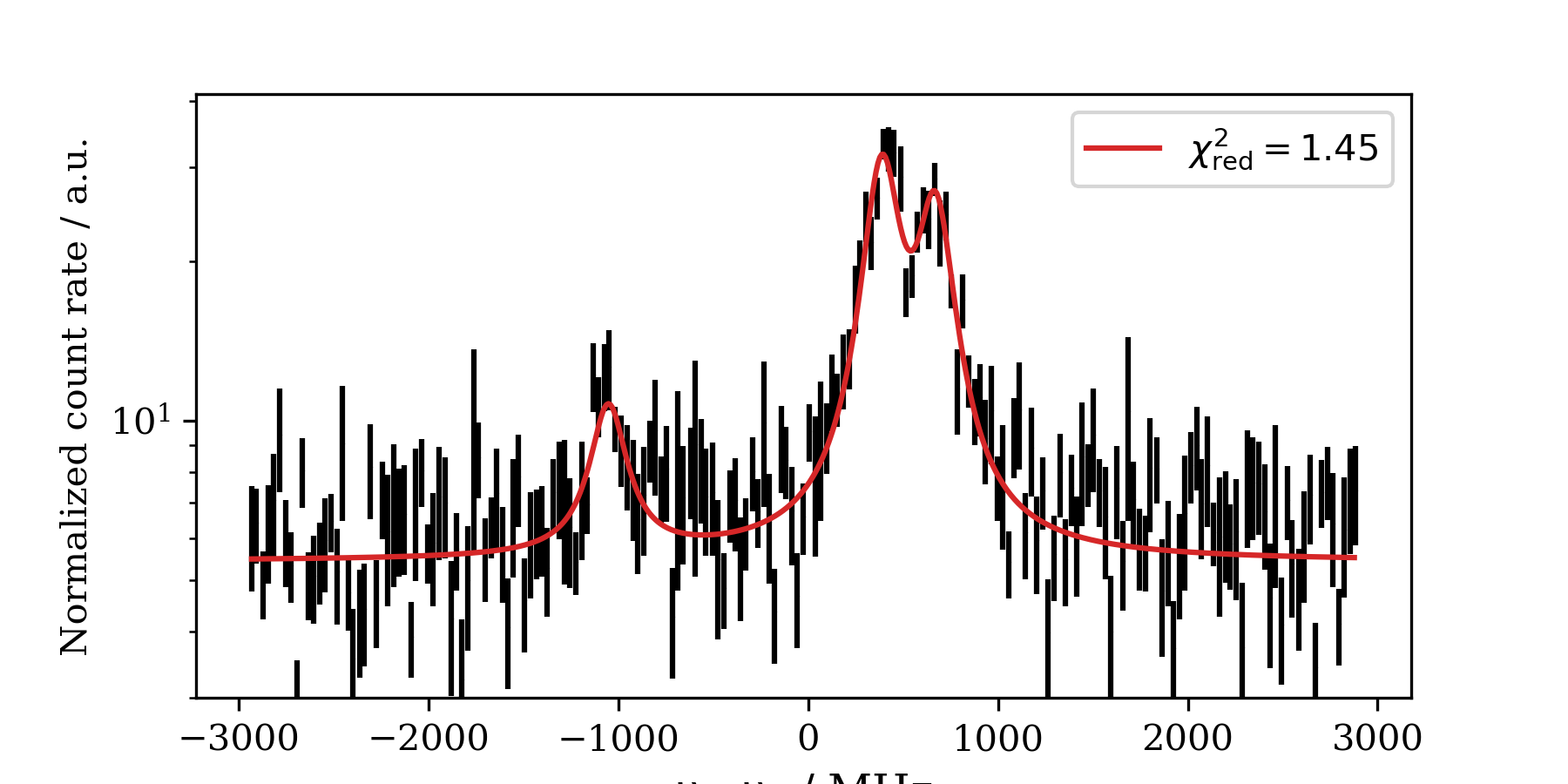}
	\caption{Measured hyperfine structure for transition B for ${}^{169}\mathrm{Tm}$. The centroid frequency of ${}^{169}\mathrm{Tm}$ stated in Fig.~\ref{fig:excitation_scheme} is taken as reference for the frequency offset $\nu-\nu_0$. The determined centroid is listed in Tab.~\ref{tab:hfs_parameters}.}
	\label{fig:tm_risiko_tran_B}
\end{figure}

\begin{figure}[htb]
	\centering \includegraphics[width= \columnwidth]{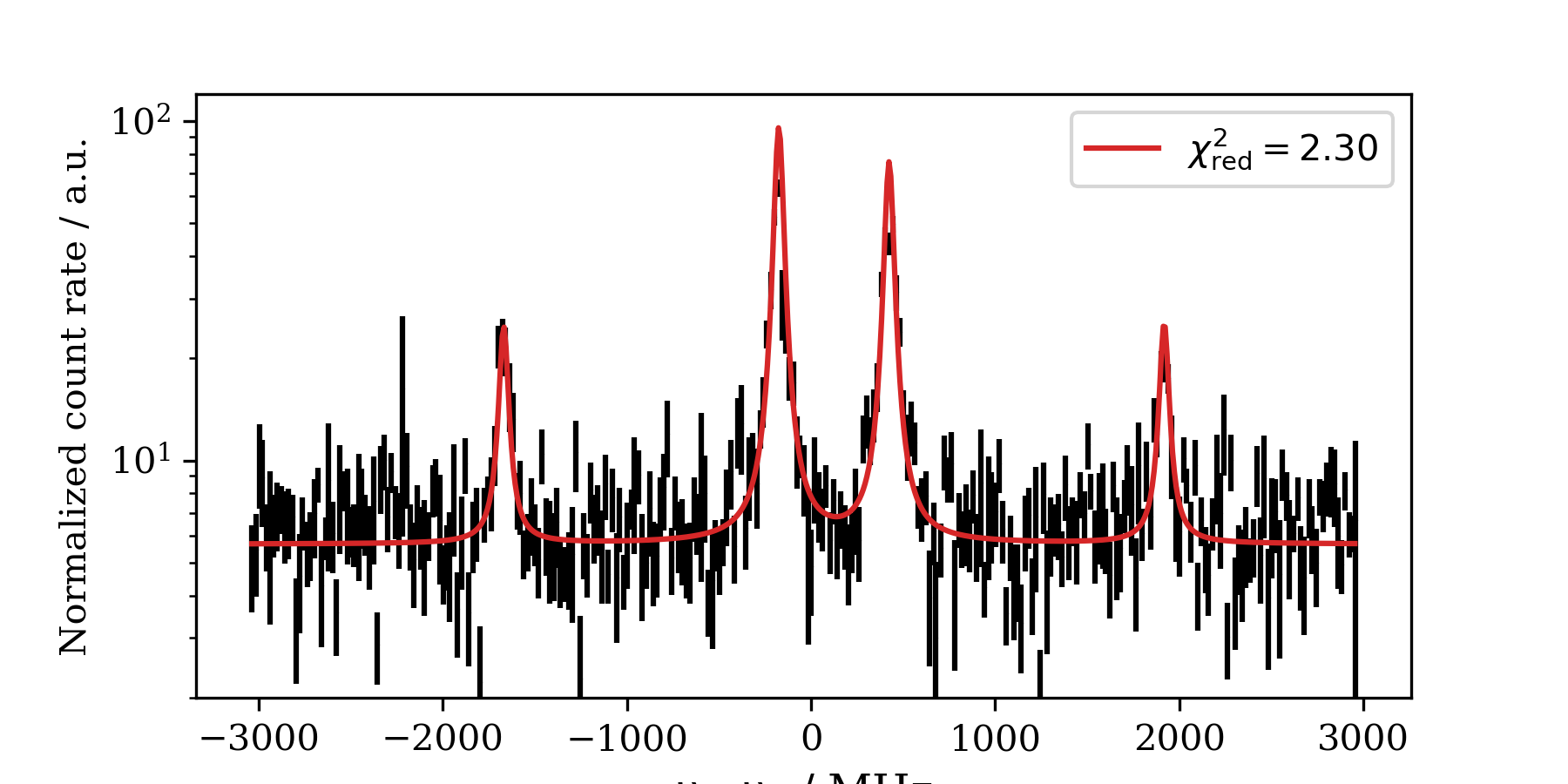}
	\caption{Measured hyperfine structure for transition C in ${}^{169}\mathrm{Tm}$. The centroid frequency of ${}^{169}\mathrm{Tm}$ stated in Fig.~\ref{fig:excitation_scheme} is taken as reference for the frequency offset $\nu-\nu_0$. The determined centroid is listed in Tab.~\ref{tab:hfs_parameters}.}
	\label{fig:tm_risiko_tran_C}
\end{figure}

The measured hyperfine structure for transition B and C for ${}^{169}\mathrm{Tm}$ are shown in Fig.~\ref{fig:tm_risiko_tran_B} and \ref{fig:tm_risiko_tran_C}. The extracted centroids from the fitting of the hyperfine structure are listed in Tab.~\ref{tab:hfs_parameters}.

\newpage
\section{Pressure-shift measurements}\label{Appendix:pressure_shift}

Complementary measurements of ${}^{169}\mathrm{Tm}$ were performed at KU Leuven to determine the pressure shift of transition A and the centroid positions of transitions B and C.
A zero-pressure reference was measured in vacuum ($\approx 10^{-7}$~\si{\milli\bar}) using an Atomic Beam Unit (ABU), while the pressure-dependent measurements were performed in an Ar-filled gas cell at the IGLIS facility~\cite{Kudryavtsev_2016,Harsh_2026}.

\begin{figure}[htb]
    \centering
    \includegraphics[width=\columnwidth]{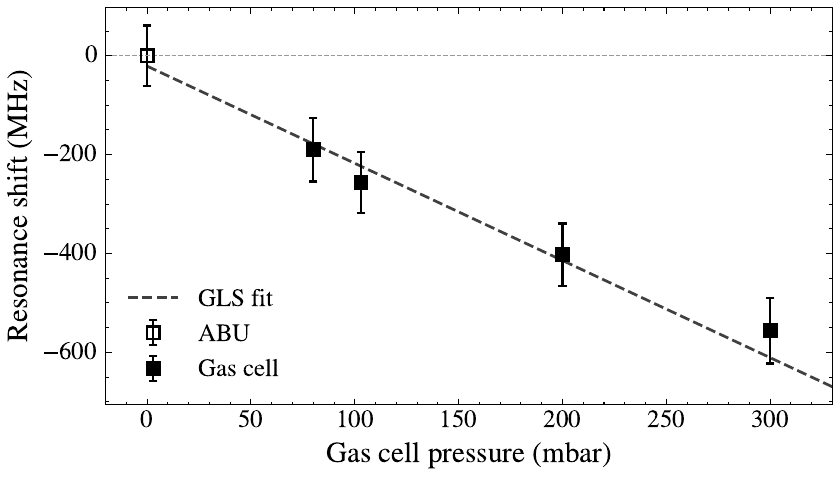}
    \caption{Measured pressure shift of transition A in ${}^{169}\mathrm{Tm}$ with Ar as buffer gas. The empty square denotes the zero-pressure ABU measurement and the dashed line the generalized-least-squares linear fit~\cite{Harsh_2026}.}
    \label{fig:pressure_shift}
\end{figure}

For transition A, the measured centroid shifts relative to the ABU value of \SI{25656.031(2)}{\per\centi\meter} are shown in Fig.~\ref{fig:pressure_shift}.
The uncertainties correspond to the quadrature sum of the statistical and systematic contributions. A generalized-least-squares linear fit yields a pressure-shift coefficient of
$\gamma_{\mathrm{shift}}=-1.99(8)_\mathrm{stat}(41)_\mathrm{sys}$~\si{\mega\hertz\per\milli\bar}. The centroid positions determined from the ABU measurements for all three transitions are summarized in Table~\ref{tab:ABU_shift}.
The statistical uncertainties were obtained from hyperfine-structure fits to the ABU spectra.
A common wavemeter uncertainty of \SI{60}{\mega\hertz} applies to all measurements.

\newpage
\begin{table}[ht]
    \centering
    \captionsetup{width=\linewidth}
    \caption{Measured centroid positions of all three transitions of ${}^{169}\mathrm{Tm}$ at the ABU experiment~\cite{Harsh_2026}. Centroid positions $\Delta \nu_i$ are relative to the values in Fig.~\ref{fig:excitation_scheme}. $\sigma_\mathrm{stat}$ is the statistical uncertainty of the fitted centroid: from a hyperfine-structure fit for the ABU measurement, and from a single Voigt profile fit for the gas-cell measurements.}
    \label{tab:ABU_shift}
    \begin{tabular}{cccc}
        \hline
        Transition & $\Delta\nu$ & $\sigma_\mathrm{stat}$ & $\tilde{\nu}_0$\\
         & [MHz] & [MHz]&[\si{\per\centi\meter}]\\
        \hline
        \hline
        A & 350 &  9.4 & 25656.019\\
        B & 350 &  1.5 & 25745.117\\
        C & 30 &  4.8 & 25717.197\\
        \hline
        \hline
    \end{tabular}
\end{table}




\end{appendices}



\end{document}